\documentclass[12pt]{iopart}

\usepackage{graphicx}
\usepackage{amssymb}

\begin{document}

\title[Percolation on hypergraphs with four-edges]{Percolation on hypergraphs with four-edges}

\author{Ojan Khatib Damavandi$^{\,1}$ and Robert M. Ziff$^{\,2}$}

\address{$^1$Department of Physics, University of Michigan, Ann Arbor, MI 48109-1040, USA\\
$^2$Center for the Study of Complex Systems 
and Department of Chemical Engineering, University of Michigan, 
Ann Arbor, MI 48109-2136, USA}
\ead{ojan@umich.edu, rziff@umich.edu}
\vspace{10pt}
\begin{indented}
\item[]August 2015
\end{indented}

\begin{abstract}
We study percolation on self-dual hypergraphs that contain hyperedges with four bounding vertices, or ``four-edges", using three different  generators, each containing bonds or sites with three distinct probabilities $p$, $r$, and $t$ connecting the four vertices.  We find explicit values of these probabilities that satisfy the self-duality conditions discussed by Bollob\'as and Riordan.  This demonstrates that explicit solutions of the self-duality conditions can be found using generators containing bonds and sites with independent probabilities.  These solutions also provide new examples of lattices where exact percolation critical points are known.  One of the generators exhibits three distinct criticality solutions ($p$, $r$, $t$).  We carry out Monte-Carlo simulations of two of the generators on two different hypergraphs to confirm the critical values.  For the case of the hypergraph and uniform generator studied by Wierman et al., we also determine the threshold $p = 0.441374\pm 0.000001$, which falls within the tight bounds that they derived.  Furthermore, we consider a generator in which all or none of the vertices can connect, and find a soluble inhomogeneous percolation system that interpolates between site percolation on the union-jack lattice and bond percolation on the square lattice.
\end{abstract}

%
%

\maketitle

\section{Introduction}
Percolation is a fundamental model in both statistical physics and mathematics  \cite{StaufferAharony94,Smirnov01,Grimmett99,BollobasRiordan06,AraujoGrassbergerKahngSchrenkZiff14,Saberi15}.  It is concerned with the formation of long-range connectivity which occurs when the occupation of sites or bonds exceeds a critical threshold.  Finding the exact threshold for different lattices is one of the primary goals in the study of percolation.  Precise thresholds are necessary for applying percolation models to real systems, and for studying the behavior of systems near the critical point.  For many years, exact values of threshold had been known for only a handful of lattices (e.g., square, triangular, honeycomb, kagome, and bow-tie lattices, for site and/or bond percolation) \cite{SykesEssam63,SykesEssam64,Wierman84}.  Recently, percolation thresholds have been found for a broad class of lattices that can be represented in the form of three-hypergraphs self-dual under the triangle-triangle transformation \cite{ziff06, wierman11}.  In a hypergraph representation, the hyperedges (in the shape of triangles, squares, etc.) can represent any collection of bonds and internal sites, including correlated sites and bonds; all that matters is the connection probabilities between the boundary vertices.   For a triangular generator, there is a single nontrivial self-duality condition which gives a unique percolation threshold  \cite{Ziff06PRE,ziff06, ChayesLei07,bollobas10}
\begin{equation}
\mathrm{Prob(all\ vertices\ connect) = Prob(none\ connect)}
\label{eq:allequalsnone}
\end{equation}
and this has led to many exactly soluble lattices when applied on self-dual three-hypergraphs \cite{ziff06}.  However, several important lattices correspond to hypergraphs with edges of four vertices (four-edges) and cannot be represented on a three-hypergraph, and it is desirable to find thresholds for such lattices as well.  Bollob\'as and Riordan \cite{bollobas10} discussed the self-duality conditions for self-dual four-hypergraphs, and recently Wierman et al.\ \cite{Wierman12} used those conditions and a stochastic ordering method to find 
bounds for a 16-bond uniform probability generator on a certain four-hypergraph.

\begin{figure}[htbp] 
    \centering
    \includegraphics[width=2in]{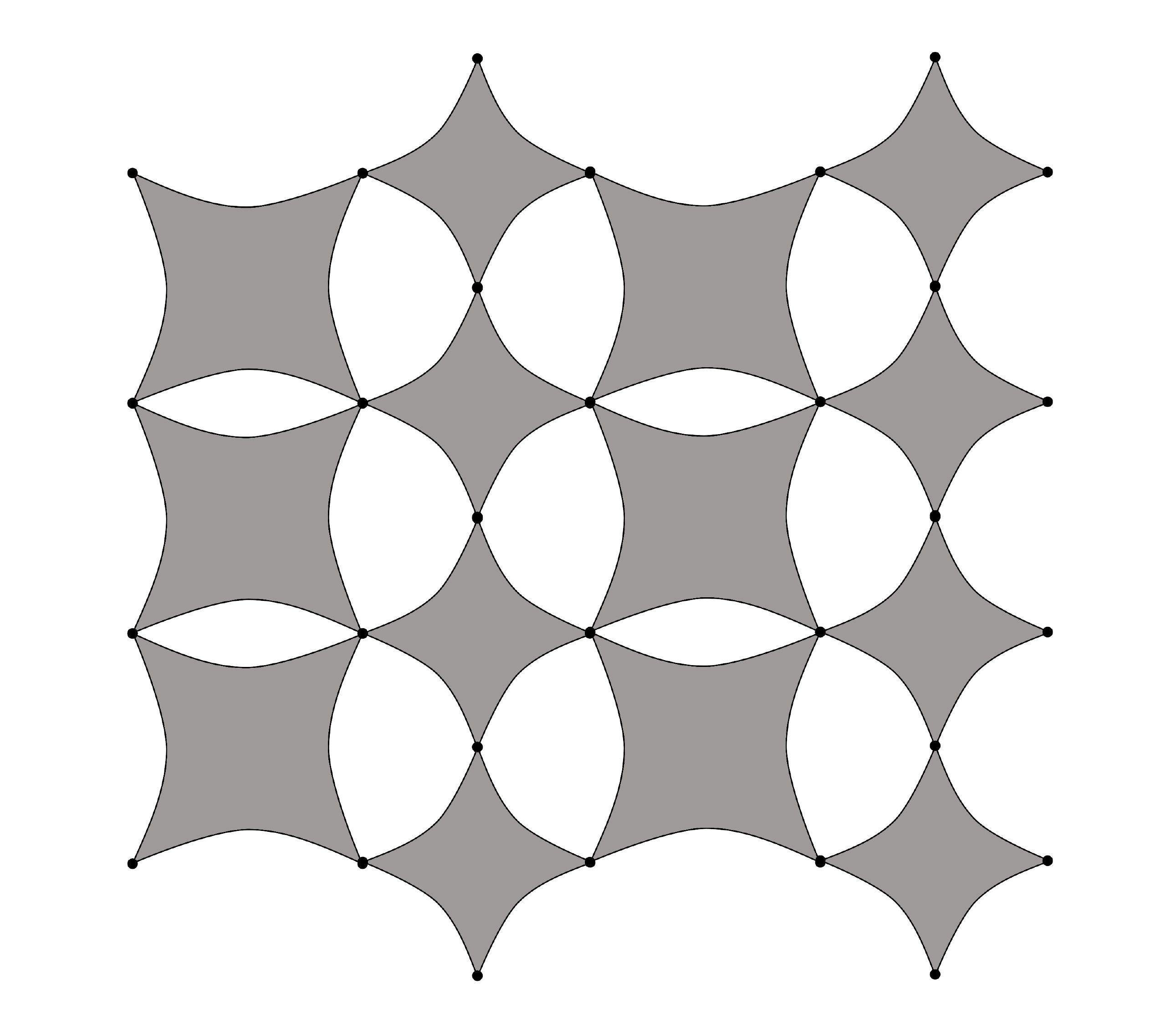} 
     \includegraphics[width=2.2in]{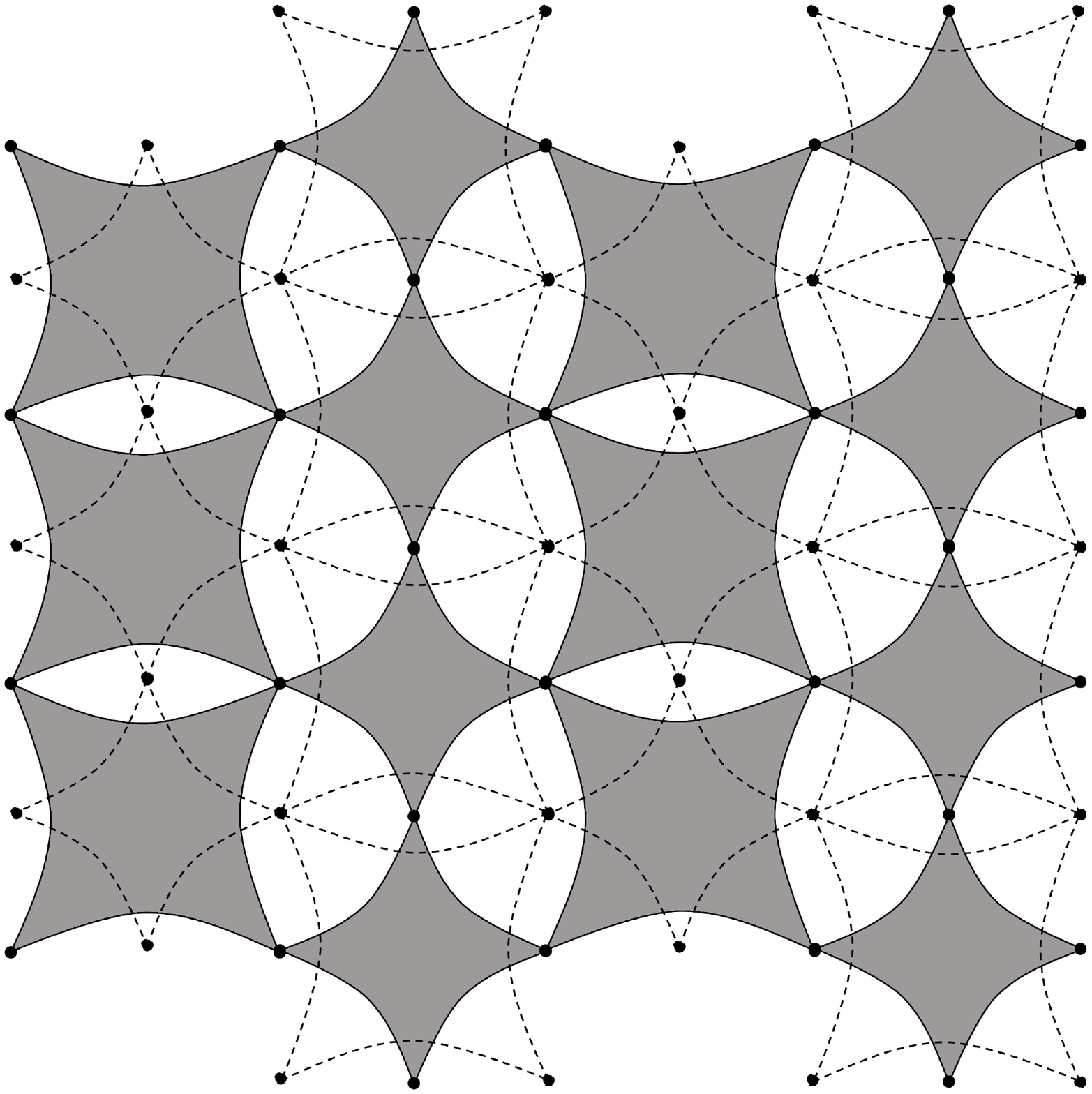} 
    \caption{(a) Hypergraph A, and (b) showing the self-duality.}
    \label{fig:hypergraphA}
 \end{figure}

\begin{figure}[htbp] 
    \centering
    \includegraphics[width=4in]{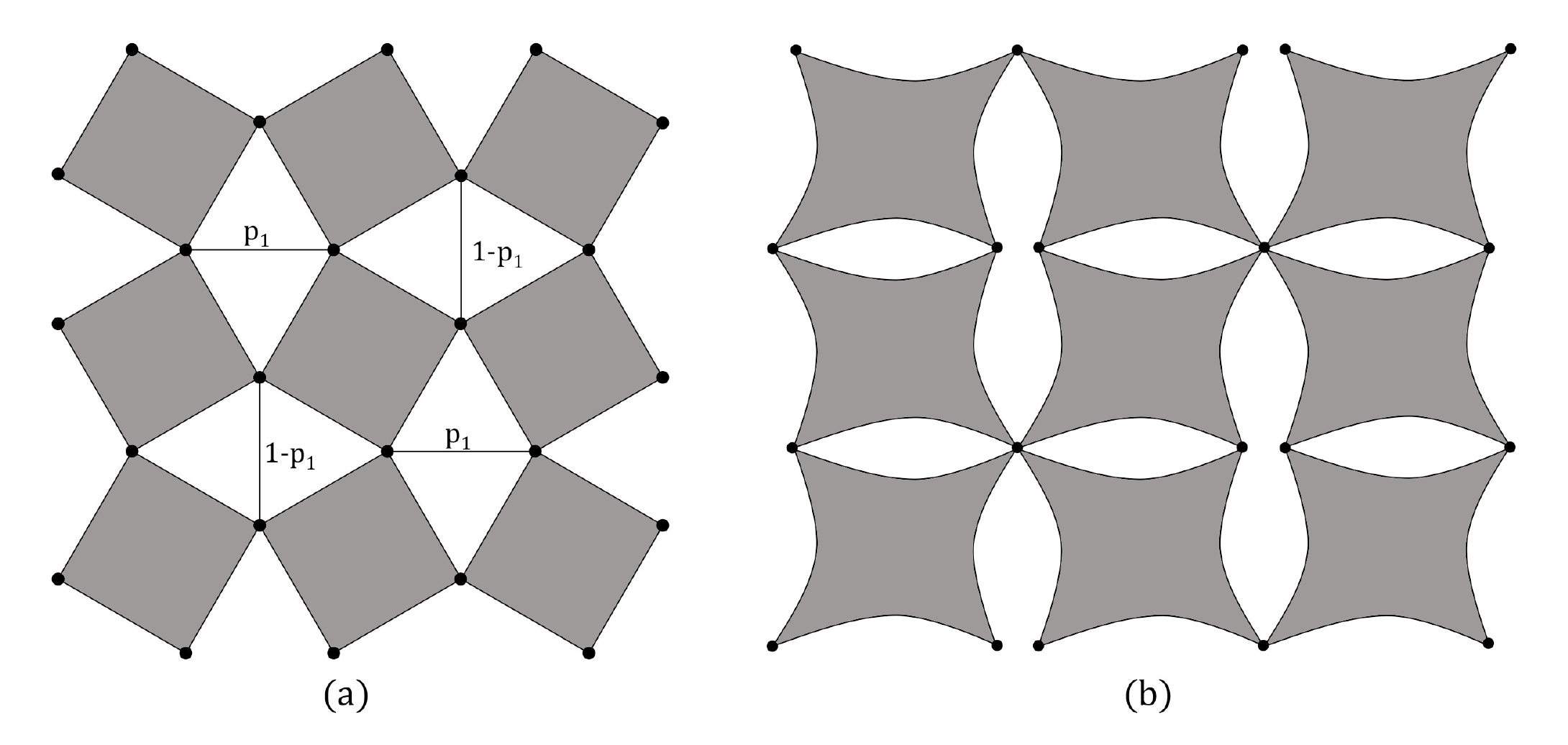} 
    \caption{(a) Hypergraph B with four-edges (squares) and single bonds of probability $p_1$ and $1-p_1$. (b) Hypergraph B with probability $p_1$ set to zero.}
    \label{fig:hypergraphB}
 \end{figure}

\begin{figure}[htbp] 
   \centering
   \includegraphics[width=4in]{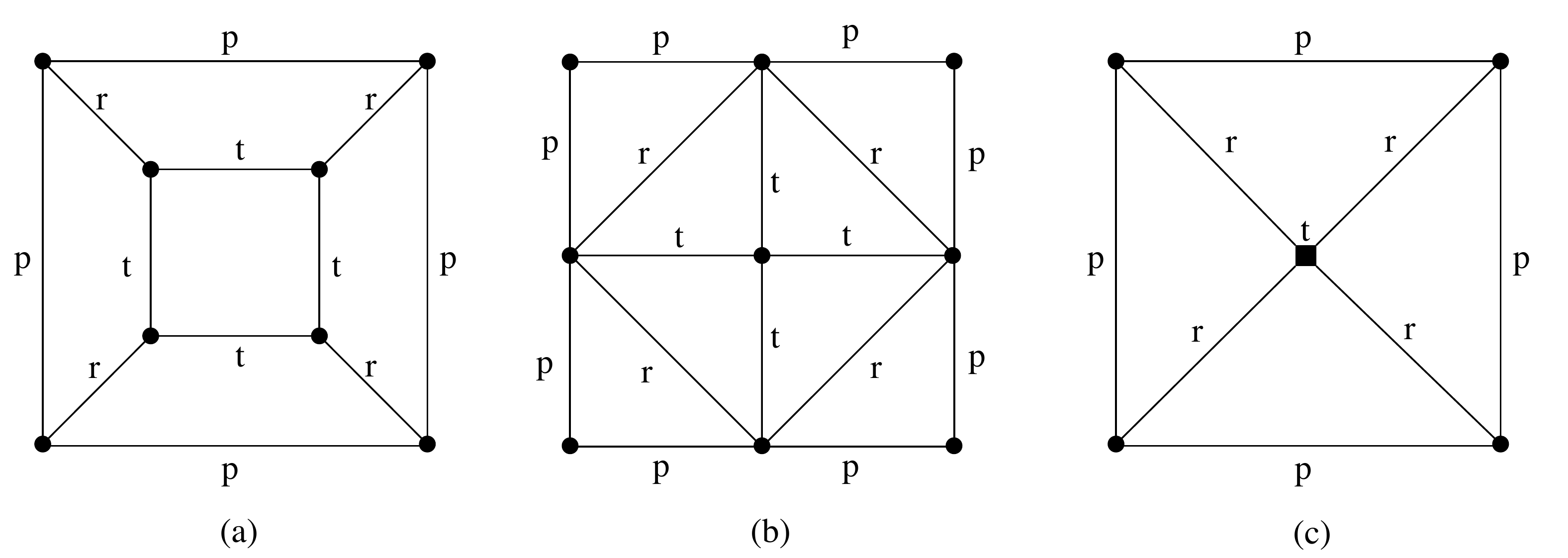} 
   \caption{(a) Generator I \space  (b) Generator II \space (c) Generator III}
   \label{fig:generators}
\end{figure}

The purpose of this paper is to find explicit self-dual four-vertex generators and to confirm with Monte-Carlo simulation that when they are placed in a self-dual hypergraph containing four-edges, the system is indeed at the threshold.   In contrast to the case of hypergraphs with triangular generators, the conditions for a square do not reduce to one single equation but to three as we will see below; therefore, we allow for three different probabilities to ensure that the set of equations have nontrivial solutions.  It will turn out that the number of solutions depends on the generator we choose.  We study the hypergraph introduced by Bollob\'as and Riordan (hypergraph A) shown in Fig.\ \ref{fig:hypergraphA} along with another hypergraph based upon the ($3^2$,4,3,4) Archimedean lattice,  which we call hypergraph B, shown in Fig.\ \ref{fig:hypergraphB}. \  We consider in detail two generators, one with 12 bonds (generator I), and one with 16 bonds introduced by Wierman et al.\ \cite{Wierman12} (generator II), both containing three distinct bond probabilities $p$, $r$ and $t$ as shown in Fig.\ \ref{fig:generators}.  We verify the criticality of these solutions numerically.  We also study the uniform case $p = r = t$ for generator II on hypergraph A numerically and find that the threshold falls well within the bounds found in \cite{Wierman12}.  Furthermore, a variant of generator I (which we call generator III) is considered by replacing the inner square in Fig.\ \ref{fig:generators}a with a single site with probability $t$ as shown in Fig.\ \ref{fig:generators}c. We do not however, perform Monte Carlo simulations for this case as this generator is not the primary focus of this paper. 
 
 Hypergraph A is a four-uniform self-dual hypergraph---i.e., consists of only four-edges connecting boundary vertices in a self-dual configuration, as shown in Fig.\ \ref{fig:hypergraphA}.  In constructing the dual, a vertex is put in each empty polygon (or face of the hypergraph), and a dual hyperedge is drawn around each hyperedge.   Hypergraph B is a bit different from hypergraph A in that it also contains two-edges (ordinary bonds). These bonds have probabilities $p_1$ and $1-p_1$ in a manner that leaves the hypergraph self-dual (Fig.\ \ref{fig:hypergraphB}a). Because the hypergraph is self-dual, the value of $p_1$ does not matter at the critical point, as long as the generators in the four-edges satisfy the duality conditions. We can therefore manipulate $p_1$ and nothing should change.  A variant of this hypergraph which only involves four-edges is obtained by setting $p_1=0$  (Fig.\ \ref{fig:hypergraphB}b). 
 
We note that another approach to finding thresholds for some lattices has recently been put forth by Grimmett and Manolescu \cite{GrimmettManolescu14}, for lattices that can be represented in an isoradial form in which each polygon can be inscribed in a circle of equal radius.  This method can be used to find a geometrical proof \cite{ZiffScullardWiermanSedlock12}
of Wu's criticality condition for the square checkerboard lattice \cite{Wu79}, however, it cannot be used to study the types of square lattices considered here.   We also note that methods have recently been developed to find approximate thresholds for many lattices to extraordinarily high precision \cite{JacobsenScullard13,Jacobsen14}.

In the following sections, we discuss the definitions (Section 2), the general self-duality conditions (Section 3), the derivation of the explicit critical points of the two generators (Section 4), Monte-Carlo simulations (Section 5), the analysis of the critical manifold for generator I (Section 6), the derivation of a generalized union-jack lattice for site percolation (Section 7), and conclusions (Section 8).  Explicit polynomials for the generators are given in the Supplementary Material.

 \section{Definitions}
 \label{defns}
Consider a generator $G$ with four boundary vertices $A$, $B$, $C$, and $D$. Denote the dual generator by $G^*$. \\
\begin{figure}[htbp] 
   \centering
   \includegraphics[width=2in]{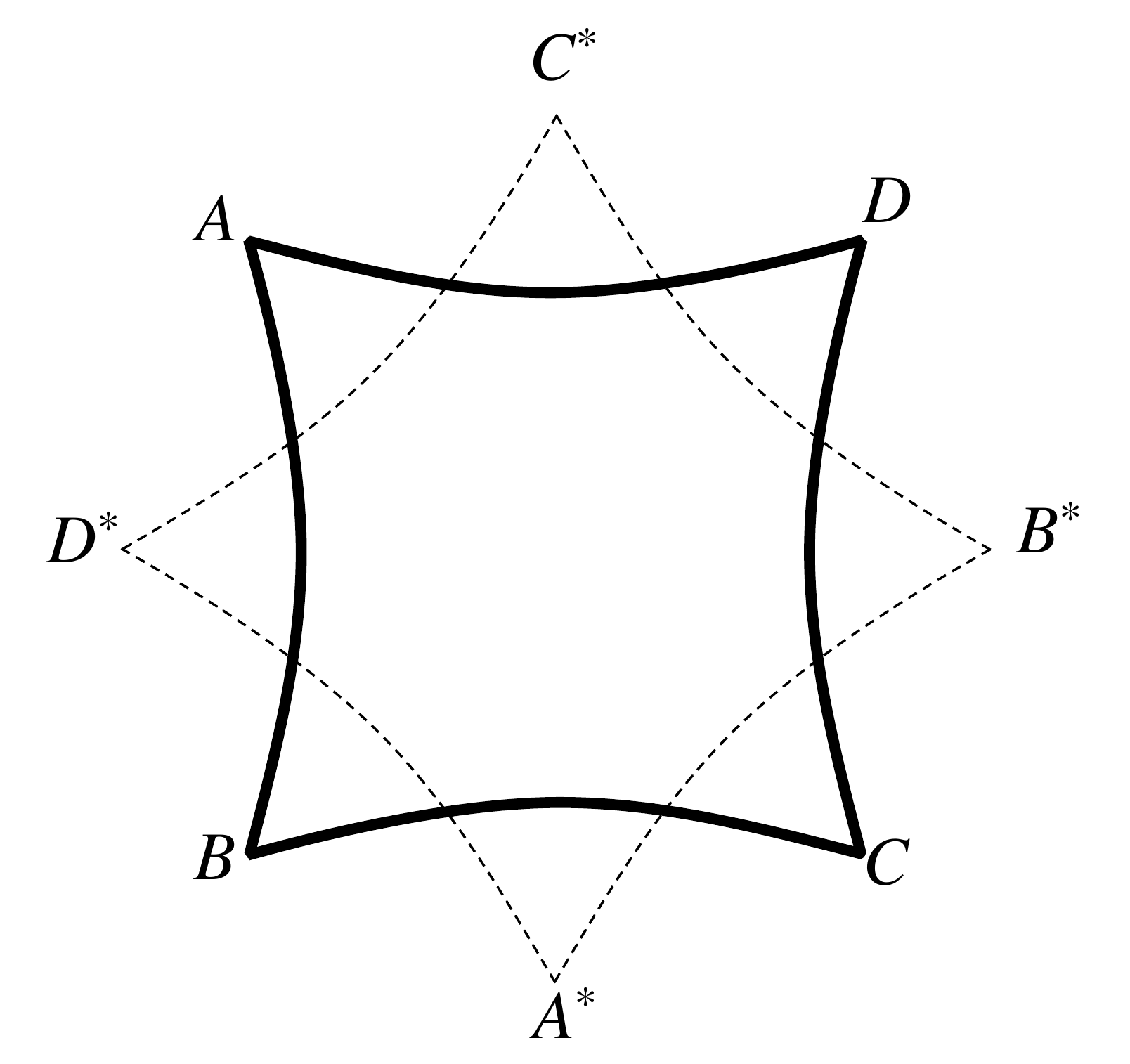} 
   \caption{Solid lines represent a four-hyperedge with boundary vertices $A$, $B$, $C$, and $D$. Dashed lines represent the dual hyperedge, with its boundary vertices $A^*$, $B^*$, $C^*$, and $D^*$.}
   \label{fig:dual}
\end{figure}
Any configuration (i.e., designation of internal edges as open or closed) on $G$ determines a partition of the boundary vertices into clusters of vertices that are connected by edges.
We use the following notation defined in \cite{wierman11}: connected vertices are grouped into clusters, and distinct  clusters are separated by a vertical bar.  For instance, $P^G [AB|CD]$ means that $A$ and $B$ are connected, $C$ and $D$ are also connected, but the two sets are not connected to each other. We also introduce the quantities $P_n$  for isotropic systems defined as follows:
\begin{eqnarray}
P_1 &=& \hbox{none of the vertices connected} = P^G [A|B|C|D] \cr
P_2 &=& \hbox{only two nearest vertices connected} \cr
 & = & P^G [AB|C|D]  \hbox{ or } P^G [BC|A|D] \hbox{ or } P^G [CD|A|B] \hbox{ or } P^G [DA|B|C] \cr
 P_3 &=& \hbox{three out of four vertices connected} \cr
 & = & P^G [ABC|D] \hbox{ or } P^G [ABD|C]\hbox{ or } P^G [ACD|B]\hbox{ or } P^G [BCD|A] \cr
P_4 &=& \hbox{all four vertices connected} = P^G [ABCD]\cr 
P_5 &=&  \hbox{only two vertices connected diagonally} \cr 
   &=& P^G [AC|B|D] \hbox{ or } P^G [BD|A|C]\cr
P_6 &=&  \hbox{two unconnected pairs} = P^G [AB|CD] \hbox{ or } P^G [AD|BC]
\label{eq:definitions}
\end{eqnarray}
Note that we cannot have $P^G [AC|BD]$; otherwise the graph will be non-planar.  Normalization requires
\begin{equation}
P_1 + 4 P_2  + 4 P_3 + P_4+ 2 P_5 + 2 P_6   = 1
\end{equation}

\section{Self-duality conditions}

\begin{figure}[htbp] 
   \centering
   \includegraphics[width=2in]{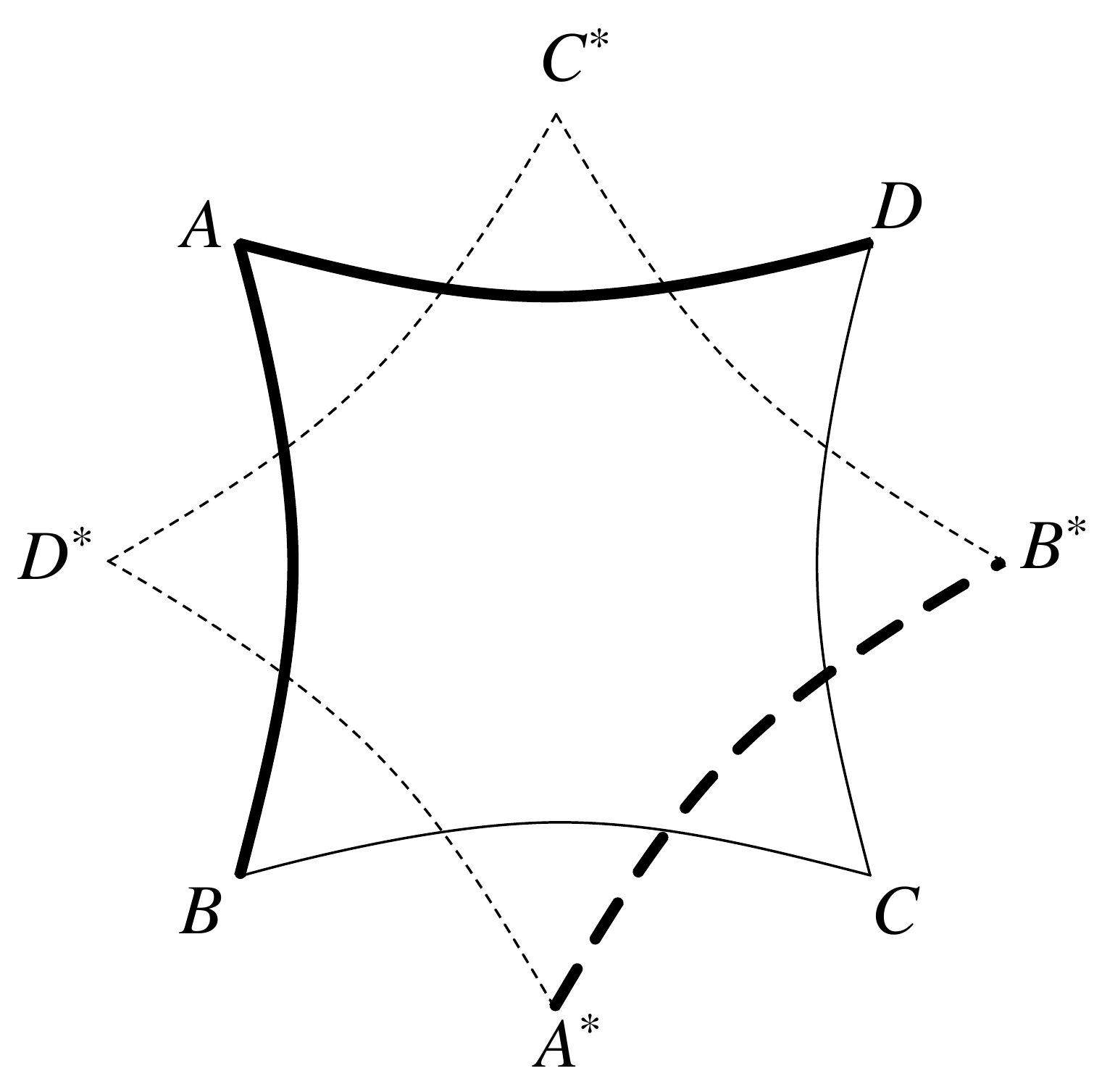} 
   \caption{If $A$, $B$, and $D$ are connected in $G$ (a case of $P_3$), only $A^*$ and $B^*$ can be connected in $G^*$ (a case of $P_2$).}
   \label{fig:duality1}
\end{figure}

\begin{figure}[htbp] 
   \centering
   \includegraphics[width=2in]{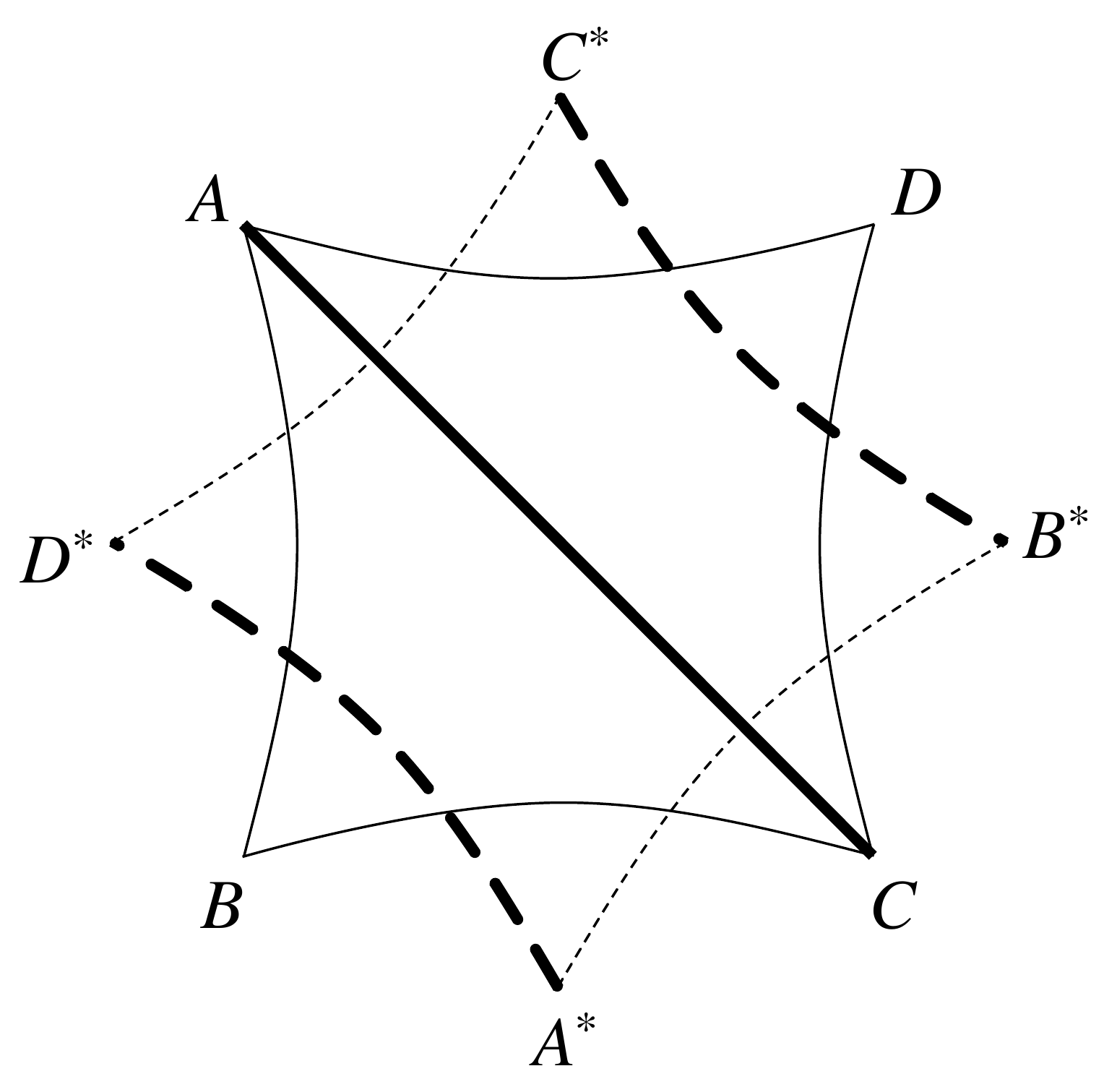} 
   \caption{If $A$ and $C$ are connected diagonally in $G$  (a case of $P_5$), then $B^*$ and $C^*$, and $A^*$ and $D^*$ can be connected in $G^*$ (a case of $P_6$).}
   \label{fig:duality2}
\end{figure}

By the duality relationship between $G$ and $G^*$, an edge in $G^*$ crossing an open edge in $G$ cannot be open. Immediately it follows that $P^G [ABCD]=P^{G^*} [A^* |B^* |C^* |D^* ]$ (Fig.\ \ref{fig:dual}). Assuming that at the critical point, $G$ and $G^*$ should have the same probability of connection between the vertices, we get the first self-duality condition which needs to be satisfied:
\begin{equation} 
P_4(p,r,t) = P_1(p,r,t)
\label{eqn1}
\end{equation}
analogous to (\ref{eq:allequalsnone}) for the triangular case.  However, here we get two additional conditions:
\begin{equation} 
P_2 (p,r,t)= P_3(p,r,t)
\label{eqn2}
\end{equation}
\begin{equation} 
P_5 (p,r,t)= P_6(p,r,t)
\label{eqn3}
\end{equation}
as given in  \cite{bollobas10}.   Figures \ref{fig:duality1} and \ref{fig:duality2} illustrate these two additional duality conditions.  The three relations yield three nontrivial equations, unlike in the case of hypergraphs with triangular generators where there is only one nontrivial equation (\ref{eq:allequalsnone} for criticality.  It is for this reason that we chose three distinct bond probabilities $(p,r,t)$ within the generators instead of only one uniform probability $p$.


\section{Derivation of the critical points} 

To find the critical points, we need explicit expressions for the various $P_n(p,r,t)$.  For the two generators, the six probabilities defined in (\ref{eq:definitions}) will be of the form
\begin{equation}
P_n(p, s, t)  =  \sum_{i=0}^{N}{\sum_{j=0}^{4}\sum_{k=0}^{4} {c^{(n)}_{ijk}p^iq^{N-i}r^js^{4-j}t^ku^{4-k}}}
\label{eq:generalform}
\end{equation}
where $q = 1 - p$, $s = 1-r$, and $u = 1 - t$, and  $N = 4$ for generator I and $N = 8$ for generator II.

To find the $c^{(n)}_{ijk}$, we used the method of exact enumeration, as follows: Go through every possibility of bond placement (there are $2^{12}$ and $2^{16}$ possibilities for generators I and II respectively); calculate the probability of percolation for each bond configuration noting that each configuration corresponds to a unique set of $\{i,j,k\}$; then count all the possible configurations that give the same probability of percolation (i.e., the same $\{i,j,k\}$); also keep track of the vertex connectivity for each configuration using standard cluster algorithms. The last two steps identify $c^{(n)}_{ijk}$ for a given $n$ and $\{i,j,k\}$.  The results of these calculations are given in the Supplementary Material.

Having found the probabilities, we then solve (\ref{eqn1}), (\ref{eqn2}) and (\ref{eqn3}) numerically to find the self-dual points. For generator II we find a single physically meaningful solution: 
\begin{eqnarray}
\label{eqn:genIIsol}
p&=&0.50003748516983960964249978886157073806 \cr
r&=&0.19560520467878219513987817879753912889 \\
t&=&0.45118455132998743413937451492248773536  \nonumber
\end{eqnarray}
while for generator I, we find three distinct solutions:\\

First solution: 
\begin{eqnarray}
\label{eqn:firstsol}
p&=&0.07878784860198622232682114813885158623 \cr
r&=&0.72008740215372047101169734956905056064\\
t&=&0.56558686100571432867919457371281967299 \nonumber
\end{eqnarray}

Second solution: 
\begin{eqnarray}
\label{eqn:secondsol}
p&=&0.14732762147606095852100839931621084702 \cr
r&=&0.59819060855262599710283661843932567345\\
t&=&0.64087266343391141007369901653243921773  \nonumber
\end{eqnarray}

Third solution: 
\begin{eqnarray}
\label{eqn:thirdsol}
p&=&0.18026397627734291307973377345301879693 \cr
r&=&0.50904978773535203861705700693678065244\\
t&=&0.75916858817391565479639455248414312467 \cr \nonumber
\end{eqnarray}
As we will discuss in Sec.\ \ref{sec:manifold}, these three points lie on a manifold in $(p,r,t)$ space. By slightly changing the values we can find the other points on the critical manifold, which now depends upon the hypergraph being considered. 
We find for a range of probabilities on the critical manifold away from these three exact solutions, the duality relations are very closely (but not exactly) satisfied. 

We also look at generator III (a variant of generator I), in which the vertices on the inner square are correlated in such a way that either all of them are connected with probability $t$ or none are connected with probability $1-t$, i.e.\ the inner square  in generator I is squeezed into a site that is open with probability $t$ or closed with probability $1-t$ (Fig.\ \ref{fig:generators}). In this case we find a single physical solution:

\begin{eqnarray}
\label{eqn:genIIIsol}
p&=&0.19738202171710149917476592797778709312 \cr
r&=&0.44960772662591992436180558214955545825\\
t&=&0.99956609784920984836488846372795998038 \cr \nonumber
\end{eqnarray}
Interestingly, the solution for $t$ is very close to 1 but not exactly 1.   If $t=1$ were a solution, then we would have a two-probability generator that satisfies the three duality relations.    Likewise, a generator with four outside bonds and two crossing diagonal bonds does not satisfy the duality conditions.  The values of $p$, $r$, and $t$ above are closest to the third solution to generator I (\ref{eqn:thirdsol}), but still quite different (especially $t$). 

For comparison, we present $P_i(p,r,t)$ for the solutions given in equations (\ref{eqn:genIIsol}--\ref{eqn:genIIIsol}) in Table~\ref{table1}. We observe that the values in any of the columns are very close to each other. In fact if we plot for example $P_i$ vs.\ $P_j$ they nearly fall on a straight line. In Sec.\ \ref{sec:manifold} we give a plot of $P_2$ vs.\ $P_1$.

\begin{table}
\caption{The values of $P_i(p,r,t)$ for the self-dual points given in (\ref{eqn:genIIsol}--\ref{eqn:genIIIsol}).}
\label{table1}
\begin{tabular*}{\textwidth}{@{}l*{15}{@{\extracolsep{0pt plus12pt}}l}}
\br
generator  & $P_1$ ($=P_4$) & $P_2$ ($=P_3$) & $P_5$ ($=P_6$) \\ 
\mr
I (soln 1) & 0.164163083    & 0.071552907    & 0.024812643    \\
I (soln 2) & 0.164308779    & 0.071547560    & 0.024750489    \\
I (soln 3) & 0.163414538    & 0.071598550    & 0.025095631    \\
II         & 0.166459728    & 0.071431975    & 0.023906186    \\
III        & 0.162628061    & 0.071642238    & 0.025401492    \\ 
\br
\end{tabular*}
\end{table}

\section{Monte-Carlo simulations}
\label{MC}

We carried out Monte-Carlo simulations using a Leath-type of growth algorithm on a lattice of size $16384 \times 16384$; clusters were grown up to a size cutoff $2^{20} = 1048576$, guaranteeing that the boundaries of the system were never reached.  This allowed us to find an unbiased estimate of $P_{\ge s} =$ the probability that a point belongs to a cluster whose size is greater than $s$.  According to scaling theory, we have at $p_c$, $P_{\ge s}  \sim s^{2 - \tau}$ where in 2d, $\tau - 2 = 5/91$, and  for $p$ close to $p_c$, we have
\begin{equation}
s ^ {\tau-2}P_{\ge s}  \sim a f(b(p-p_c) s^\sigma) \approx A + B (p - p_c) s^\sigma 
\label{eq:Pscaling}
\end{equation}
where $f(z)$ is the universal scaling function, and the last term above follows from a Taylor series expansion of $f(z)$, where $a$, $b$, $A$, and $B$ are non-universal constants which are specific to the system being studied, while $\sigma=36/91$ and $f(z)$ are universal.  Thus a plot of $s ^ {\tau-2}P_{\ge s}$ vs.\ $(p - p_c) s^\sigma$ yields a straight line (for large $s$) with a slope proportional to $p-p_c$.  When $p = p_c$, $s ^ {\tau-2}P_{\ge s}$ is constant, apart from deviations for small $s$ where scaling is not valid.   

\subsection{Generator I on hypergraph A} 
\begin{figure}[htbp] 
   \centering
   \includegraphics[width=1in]{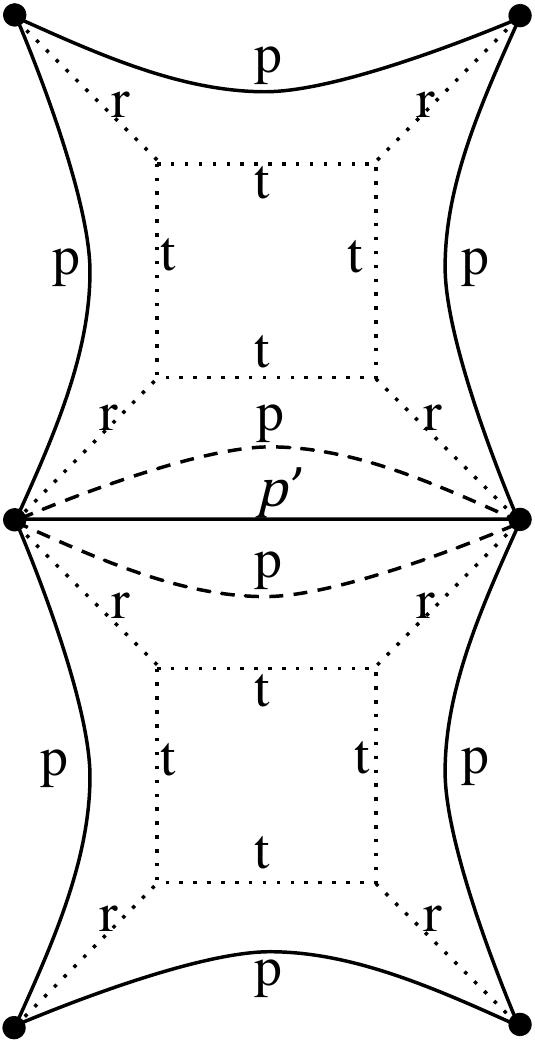} 
   \caption{Replacing the two central parallel bonds (dashed lines) having probability $p$ by one bond having effective probability $p'=2p-p^2$, necessary when Generator I on hypergraph A is used. The dotted lines represent the interior of the generator with probabilities defined in Fig.\ \ref{fig:generators}a.}
   \label{fig:replacebond}
\end{figure}

\begin{figure}[htbp] 
   \centering
   \includegraphics[width=4in]{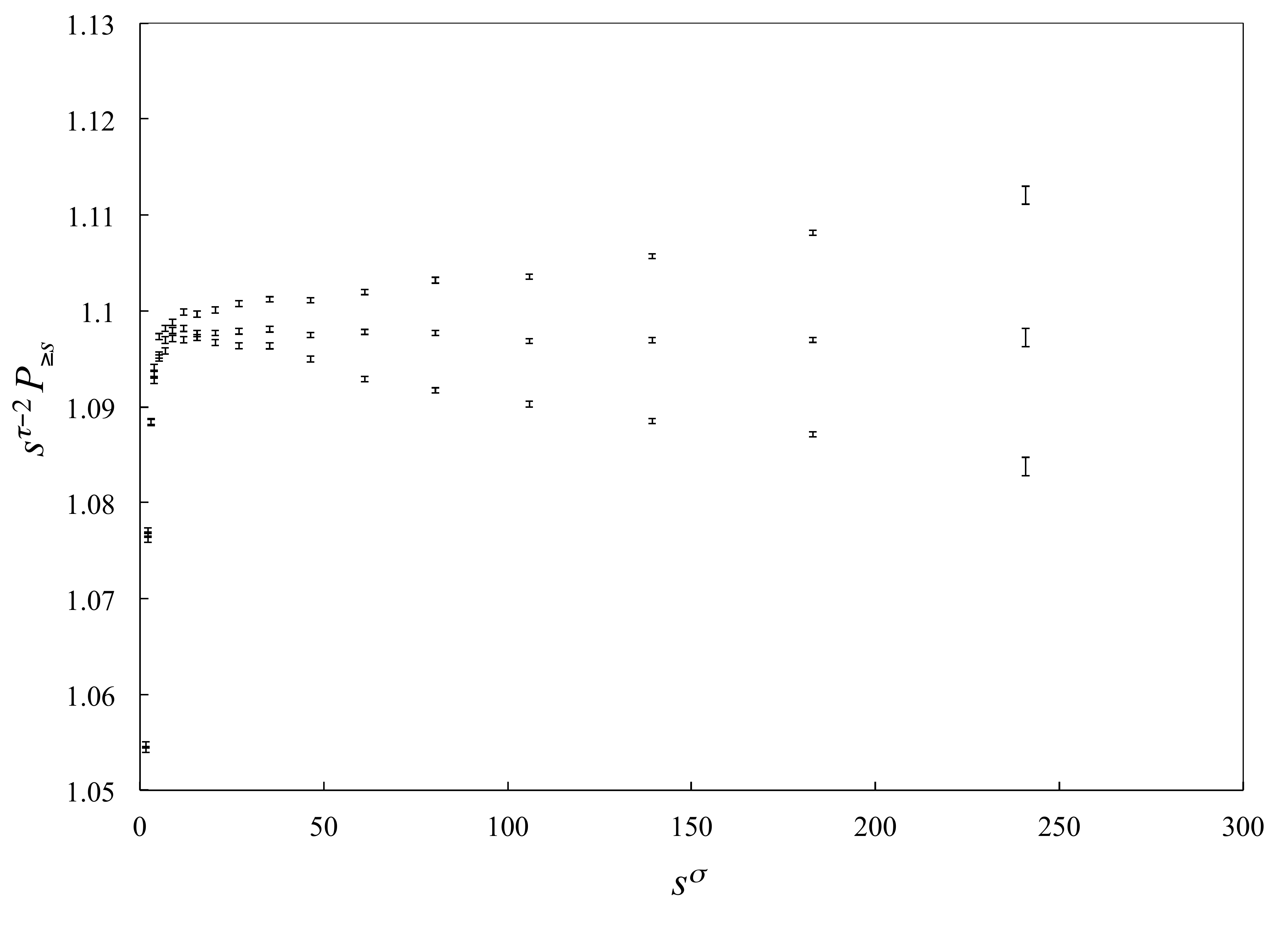} 
    \caption{Plot of $s^{\tau-2} P_{\geq{s}} $ vs.\ $s^\sigma$ for the Monte-Carlo simulation of generator I on hypergraph A, for the first self-dual point, eqn.\ (\ref{eqn:firstsol}) (middle),  for $p+0.0001$ and the same $r$ and $t$ as in (\ref{eqn:firstsol}) (upper), and for $p-0.0001$ and again the same $r$ and $t$ (lower).  The horizontal plot for the first case confirms that the system is at the critical point with those probabilities. }     
\label{fig:MonteCarlo}
\end{figure}

In this case, because we have two parallel bonds where the square generators touch, it was convenient to replace those two bonds having probability $p$ by one bond having probability $p'=2p-p^2$  \cite{Wierman84}, as shown in Fig.\ \ref{fig:replacebond}.  Figure \ref{fig:MonteCarlo} shows the results for the simulation of 250\,000 samples on a lattice of $16384 \times 16384$ for generator I on hypergraph A, at the predicted critical point of the first solution (\ref{eqn:firstsol}) and with values of $p$ equal to $0.0001$ above and below the critical value $0.07878785\ldots$.  It can be seen that (\ref{eqn:firstsol}) corresponds to a critical point within high numerical accuracy (at least $10^{-5}$).  Likewise, we verified that the other two solutions are also at the critical point within this error.

\subsection{Generator II on hypergraph A} 
We also confirmed the criticality of the self-dual point (\ref{eqn:genIIsol}) of generator II on hypergraph A by a Monte-Carlo simulation of 150\,000 samples, and find results similar to Fig.\ \ref{fig:MonteCarlo} but will not be shown here.  In addition, we carried out extensive simulations for the case of uniform probabilities with this generator and hypergraph, and found the critical value $p=r=t=0.441374\pm 0.000001$, which falls well within the bounds given by Wierman et al.\  \cite{Wierman12}:\footnote{We mention that there appears to be a minor calculation error in \cite{Wierman12}.  In section 4.2, which concerns the model considered here, the authors give the fourth probability as 0.432569051787763.  However, using Mathematica to solve the fifth equation in their Table~4,  $P_p[C_1] + P_p[C_2] + P_p[C_3] = P_p[C_4] + P_p[C_5] + P_p[C_6]$, we find instead $p = 0.4444066898118085$.  (The fifth probability they give is the solution to the fourth equation in Table~4.)  However, because this value falls within their other values, it does not alter the final bounds (\ref{eq:wiermanbounds}).}  
\begin{equation}
0.424072 \le p_c \le 0.463661
\label{eq:wiermanbounds}
\end{equation}
This is a non-self-dual percolation threshold, specific to the uniform generator II on hypergraph A.  At this value of $p$, the connection probabilities are $P_1 = 0.174841$, $P_2 = 0.0645187 $, $P_3 = 0.0786097$, $P_4 = 0.153594$, $P_5 = 0.0375026$, and $P_6 = 0.0120231$.  The duality conditions  (\ref{eqn1}--\ref{eqn3}) are far from being satisfied by these values, and the values of the $P_i$ are far from those of the other generators listed in Table~\ref{table1}.   

\begin{figure}[htbp] 
   \centering
   \includegraphics[width=4in]{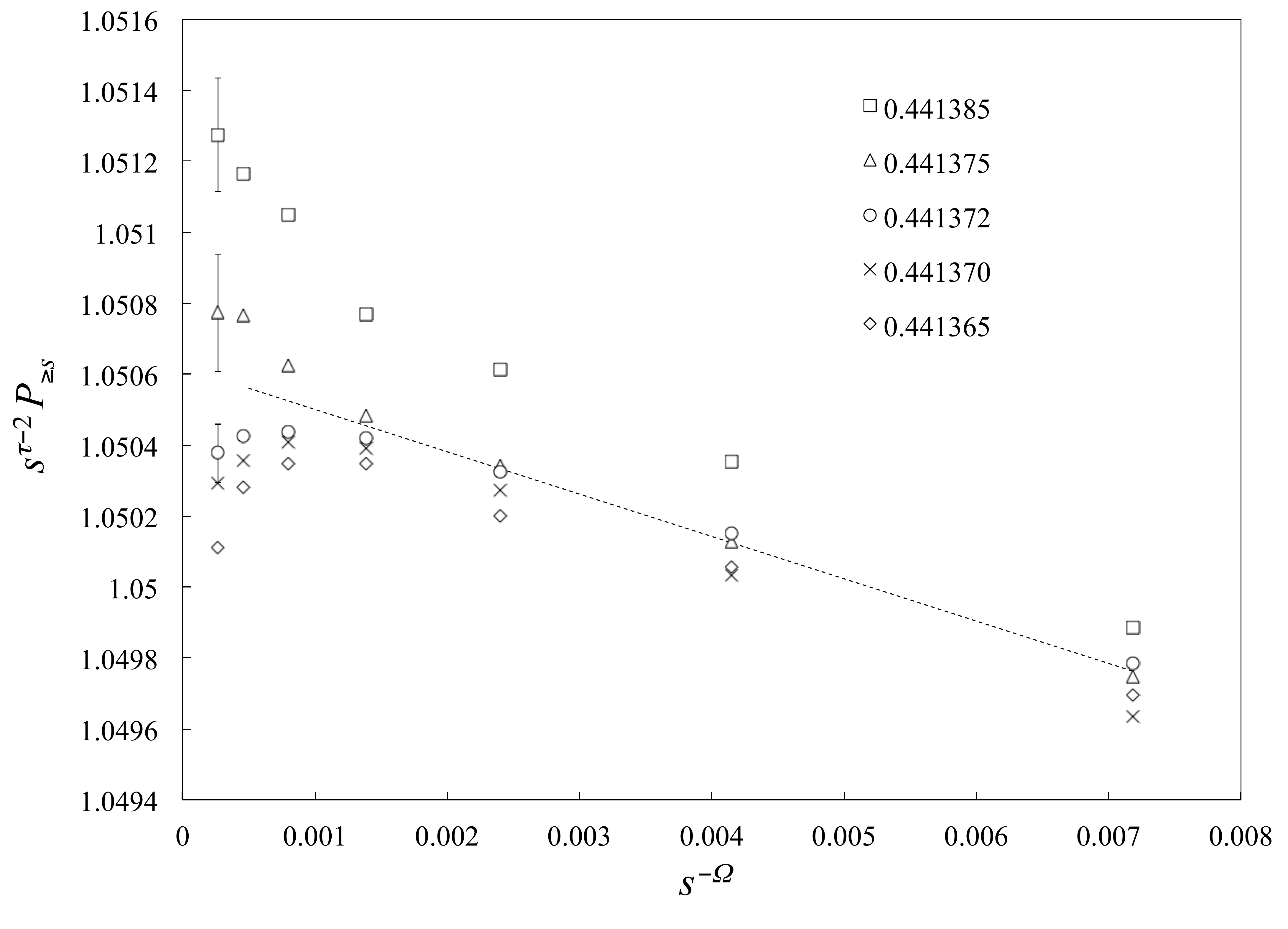} 
   \caption{Plot of $s^{\tau-2}P_{\ge s}$ vs.\ $s^{-\Omega}$ with $\Omega = 72/91$ for Monte-Carlo results of the uniform probability generator II $p = r = t$ on hypergraph A.  The dashed line represents the linear behavior which is expected for $p = p_c$, and suggests $p_c = 0.441374 \pm 0.000001$.  Some representative error bars are shown.}
   \label{fig:wiermanuniform}
  \end{figure}

To find this result we simulated up to $10^8$ samples for each value of $p$ using a smaller cutoff of $2^{15} = 32768$.  Our Monte-Carlo results are plotted in Fig.\ \ref{fig:wiermanuniform}.  Here we show an alternate way of analyzing the data, where we plot $s^{\tau-2}P_{\ge s}$ vs.\ $s^{-\Omega}$ where $\Omega = 72/91$ \cite{Ziff11,AharonyAsikainen03} represents the finite-size scaling for smaller $s$.  At $p_c$, we have  $s^{\tau-2}P_{\ge s} = A + C s^{-\Omega}$, so at $p_c$ our plot should yield a straight line.  This form of plotting the data is useful when $p$ is very close to $p_c$.

\subsection{Generator I on hypergraph B} 
Here, we have the two additional bonds, one with probability $p_1$ and the other with probability $p_2=1-p_1$. However, since we are looking at self-dual square generators, which do not depend on $p_1$, we expect that by changing $p_1$ nothing changes, and this is exactly what is observed.  We performed the simulation for four different values of $p_1$ (0.00, 0.30, 0.50, and 0.85) and for the three self-dual points given in equations (\ref{eqn:firstsol}--\ref{eqn:thirdsol}), the system was found to be critical in all of these cases within numerical accuracy.

\subsection{Generator II on hypergraph B} 
We verified that the self-dual point (\ref{eqn:genIIsol}), is the critical point for the four values of $p_1$ used above, consistent with our expectations. We also tested the uniform probability critical point ($p = r = t = 0.441374$) that we found for generator II hypergraph A, and observed that in this case, as expected, $p_1$ does matter.  The system is at criticality only for the two points $p_1\approx 0.7665$ and $p_1 \approx 1 - 0.7665 =  0.2335$.

\section{Further analysis of critical manifolds}
\label{sec:manifold}
\begin{figure}[htbp] 
   \centering
   \includegraphics[width=3in]{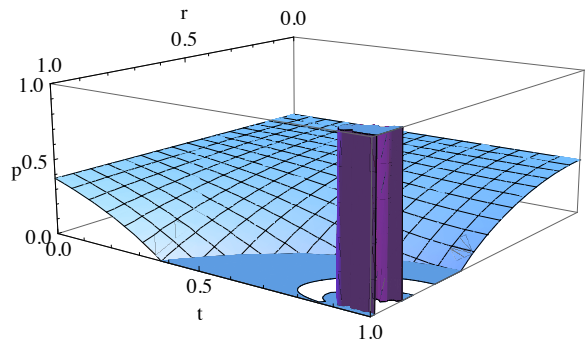} 
   \caption{$p^{(1)}(r,t)$ satisfying $P_4=P_1$. The column at the corner of the plot is the artifact of the plotting program caused by a singularity.}
   \label{fig:P4P0}
 \end{figure}

\begin{figure}
   \centering
   \includegraphics[width=3in]{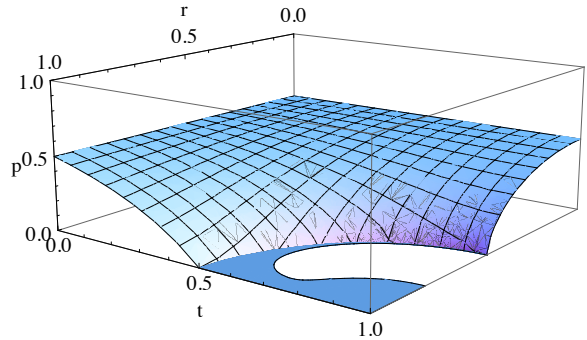} 
   \caption{$p^{(2)}(r,t)$ satisfying $P_2=P_3$.}
   \label{fig:P2P3}
 \end{figure}

\begin{figure}
   \centering
   \includegraphics[width=3in]{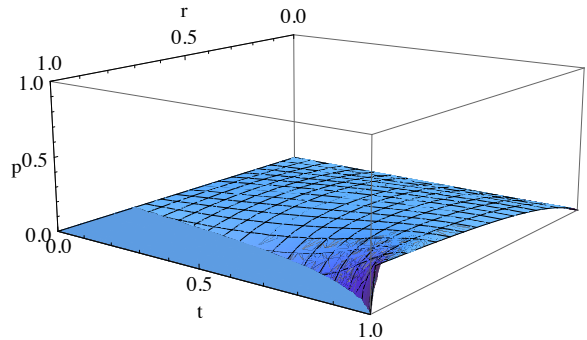} 
   \caption{$p^{(3)}(r,t)$ satisfying $P_5=P_6$.}
   \label{fig:P5P6}
\end{figure}

For generator I, we further studied the self-dual relations. We independently evaluated each of the three conditions given in eqns.\ (\ref{eqn1}), (\ref{eqn2}) and (\ref{eqn3}) and found $p$ in terms of $r$ and $t$. In this way, we found three equations for $p(r,t)$, one for each duality condition:
\begin{equation}
P_4(p,r,t) = P_1(p,r,t) \Rightarrow p^{(1)}(r,t)
\end{equation}
\begin{equation}
P_2(p,r,t) = P_3(p,r,t) \Rightarrow p^{(2)}(r,t)
\end{equation}
\begin{equation}
P_5(p,r,t) = P_6(p,r,t) \Rightarrow p^{(3)}(r,t)
\end{equation}
Figures \ref{fig:P4P0}, \ref{fig:P2P3} and \ref{fig:P5P6} show the manifolds for the three self-duality conditions. These three manifolds intersect at three points---precisely our self-dual points. \\


\begin{figure}[htbp] 
   \centering
   \includegraphics[width=3in]{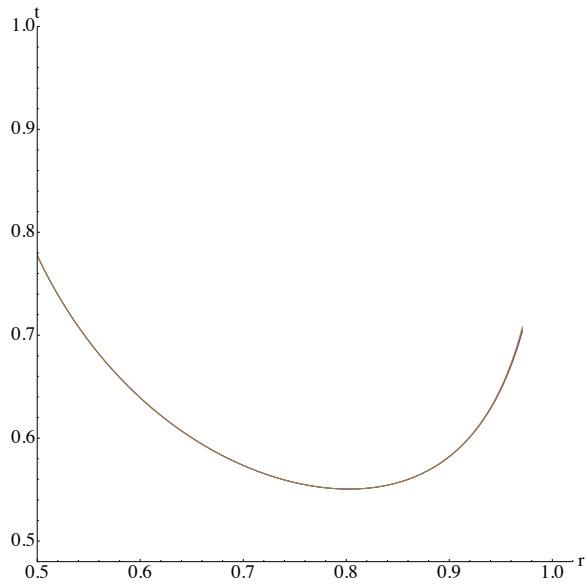} 
   \caption{Plot of the curves of $t^{(1)}(r)$, $t^{(2)}(r)$ and $t^{(3)}(r)$. There are actually three different curves here but they are not distinguishable in this figure.}
   \label{fig:contour}
\end{figure}

\begin{figure}[htbp] 
   \centering
    \includegraphics[width=4in]{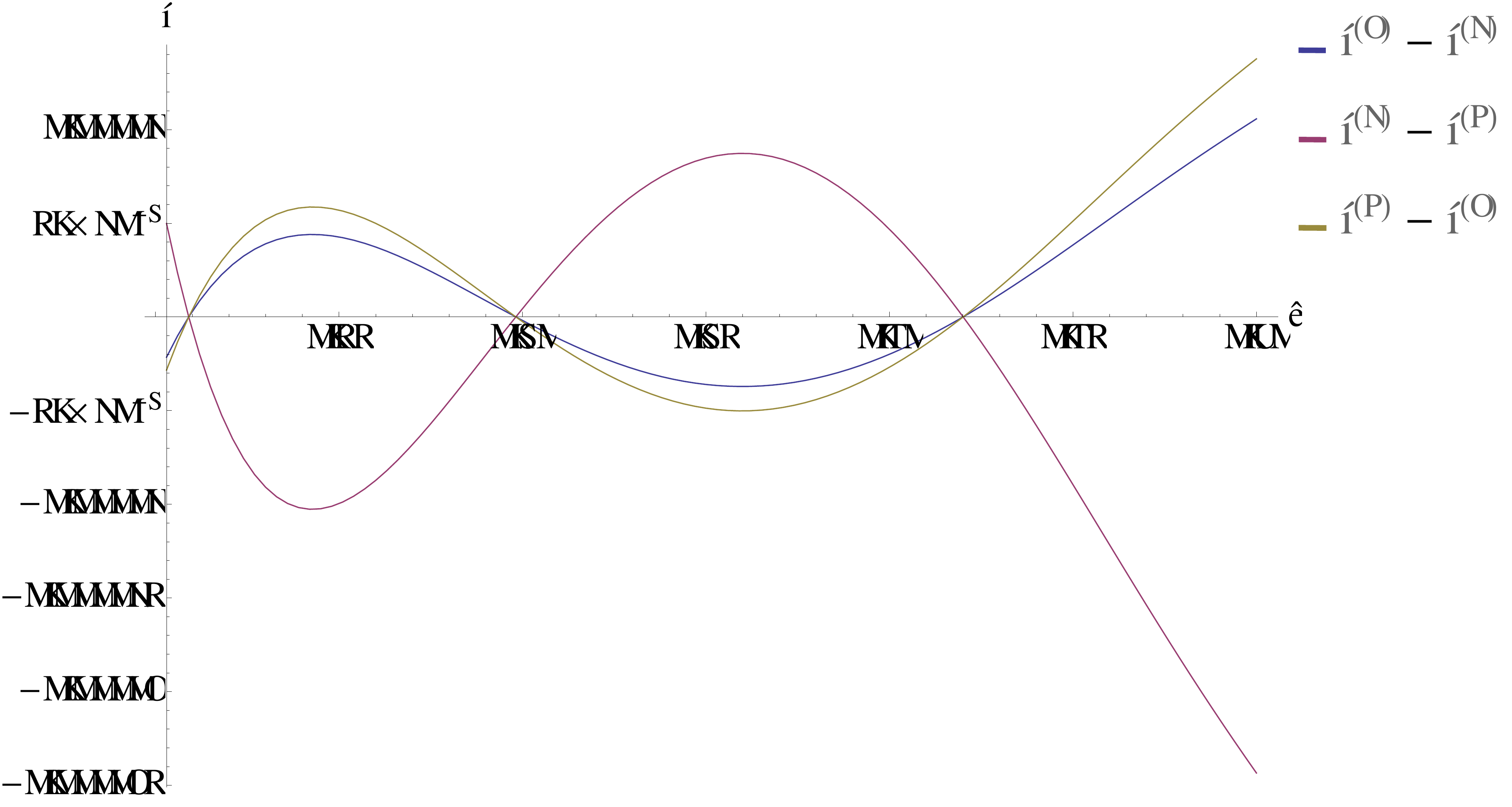} 
   \caption{Plot of the difference between pairs of curves of $t^{(n)}(r)$ of Fig.\ \ref{fig:contour}. 
   The curves cross exactly at self-dual points (\ref{eqn:firstsol}--\ref{eqn:thirdsol}).}
   \label{fig:r-t}
\end{figure}

To see the intersection of these surfaces, we set pairs of the $p^{(n)}(r,t)$ equations equal to each other and find $t$ as a function of $r$. This gives us three non-independent relations $t(r)$:
\begin{equation}
p^{(1)}(r,t) =  p^{(2)}(r,t) \Rightarrow t^{(1)}(r)
\label{t-first}
\end{equation}
\begin{equation}
p^{(1)}(r,t) =  p^{(3)}(r,t) \Rightarrow t^{(2)}(r)
\end{equation}
\begin{equation}
p^{(2)}(r,t) =  p^{(3)}(r,t) \Rightarrow t^{(3)}(r)
\end{equation} 
where $t^{(1)}(r)$ is satisfied if both self-duality conditions (\ref{eqn1}) and (\ref{eqn2}) are satisfied, and so forth. The interesting fact is that if we plot  the curves of $t^{(1)}(r)$, $t^{(2)}(r)$ and $t^{(3)}(r)$, they are very nearly identical in a wide region (Fig.\ \ref{fig:contour}), which implies that the points along these curves should be quite close to self-duality.
Finally, we plot their differences in Fig.\  \ref{fig:r-t}, which shows clearly that the curves cross at the three self-dual points.  The curves remain within $0.00001$ of each other over a wide range of $r$, meaning that we ``almost" have a manifold of thresholds. 

\begin{figure}[htbp] 
   \centering
   \includegraphics[width=4in]{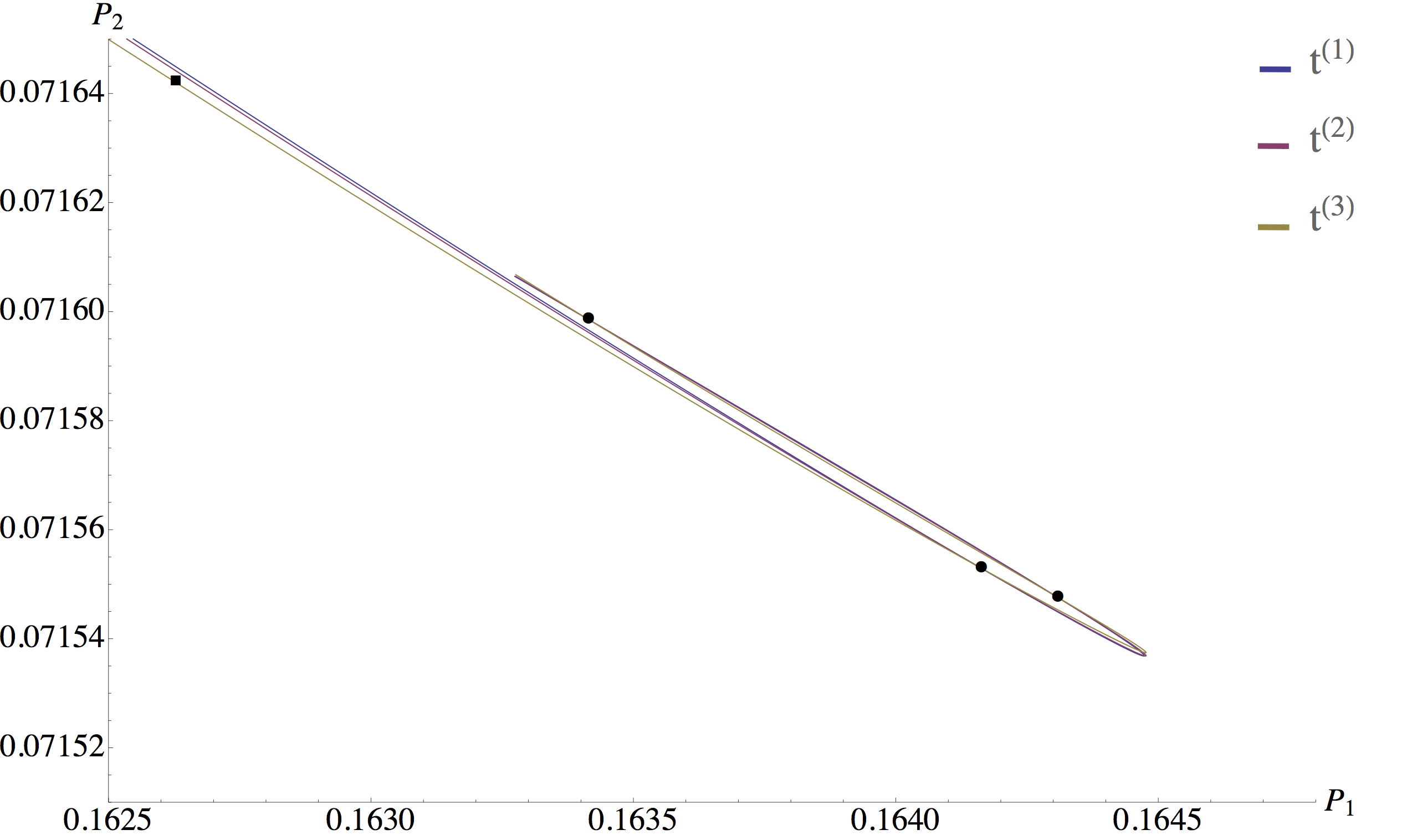} 
   \caption{Plot of $P_2(p,r,t)$ vs.\ $P_1(p,r,t)$ for the points $t^{(i)}(r)$ along the curves in Fig.\ \ref{fig:contour}. The circles mark the values of the self-dual probabilities. These are indeed at the crossing points of the curves. The black square shows the probabilities for generator III, which falls nearly on the curve determined by $t^{(3)}(r)$.}  
\label{fig:P2vP1}
\end{figure}

Motivated by the results given in Table~\ref{table1}, we find $P_i(p,r,t)$ along the curves in Fig.\ \ref{fig:contour}. $P_2$ vs.\ $P_1$ is plotted in Fig.\ \ref{fig:P2vP1}. Also on the same plot, we show the values for the self-dual points of generator I (where the three curves cross) along with generator III, which is a variant of generator I. It can be seen that the points lie almost on a straight line; we have no explanation for this surprising result.  The point $(P_1,P_2)$ for generator II falls on the same approximate line, however, it is not shown on Fig.\ \ref{fig:P2vP1} as it is farther away from other points on the plot. Similar nearly linear curves can be obtained for other $P_i$ vs.\ $P_j$.

The same analysis was not possible for generator II because the $p^{(n)}(r,t)$'s were found to be  singular at many points, and solutions for $t^{(n)}(r)$ could not be found.

\section{Results for site percolation}
\begin{figure}[htbp] 
   \centering
   \includegraphics[width=2in]{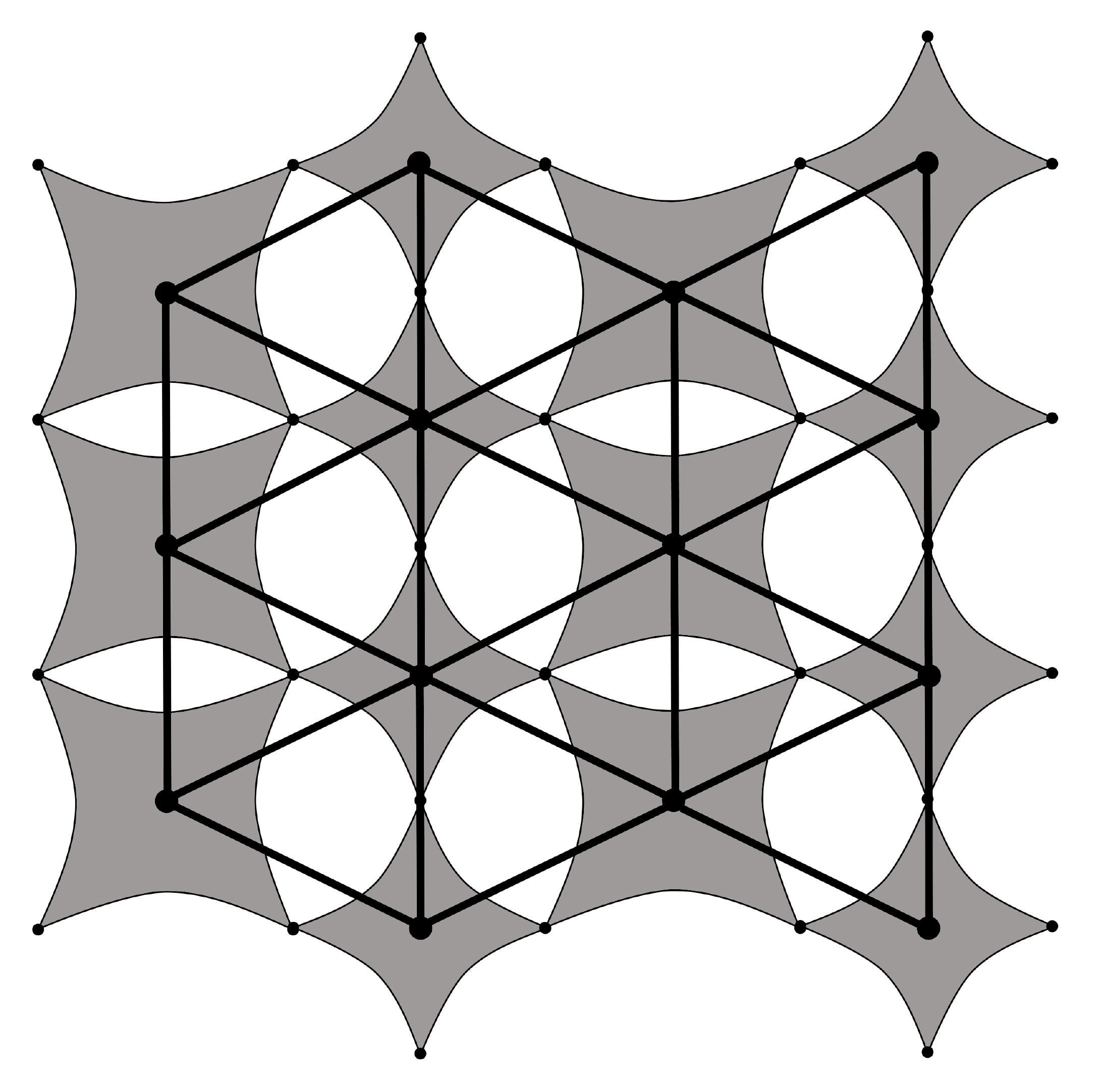} 
   \caption{If each hyperedge is replaced by a site in hypergraph A, we get a triangular site percolation system.}
   \label{fig:triangular-site}
   \includegraphics[width=2in]{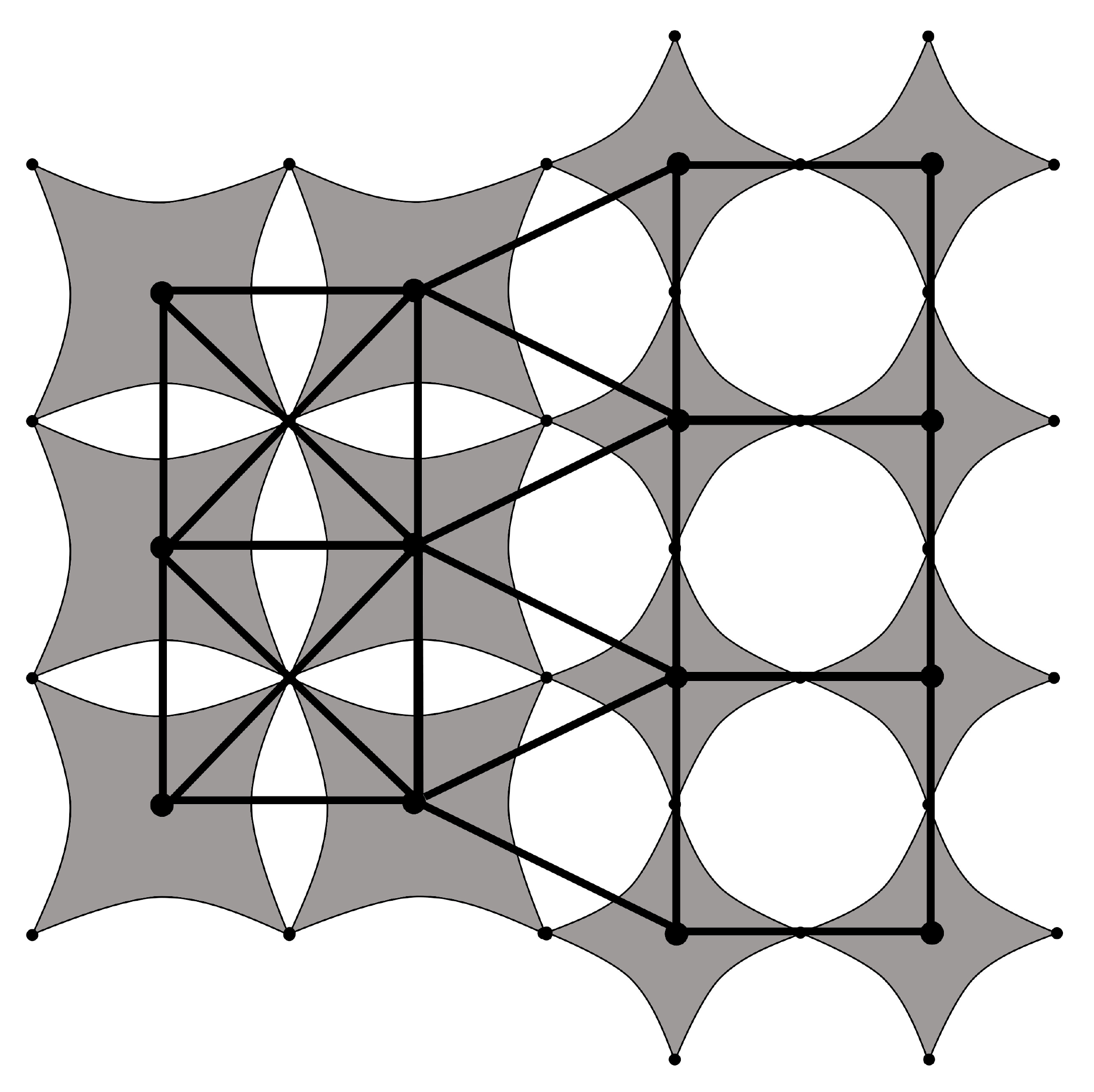}
   \caption{Replacing all hyperedges by a site in this self-dual hypergraph created by ``stretching out" hypergraph A yields another system with site percolation threshold of $p = 1/2$, but with non-planar crossing bonds (on the left).  The average coordination number of the site lattice is six. }
    \label{fig:newdualconfig}
\end{figure}

By choosing a generator in which either all or none of the four vertices are connected together, we can generate critical site percolation systems.  (For the three-hypergraph case where the hypergraph is a regular array of triangles, this procedure yields simply the system of site percolation on the triangular lattice.)  Thus, we assume $P_2 = P_3 = P_5 = P_6 = 0$, and the self-duality conditions are satisfied if $P_1 = P_4 = 1/2$.   Applying this to hypergraph A, we find site percolation on the triangular lattice once again (Fig.\ \ref{fig:triangular-site}).  In Fig.\ \ref{fig:newdualconfig} we consider a stretched version of hypergraph A, which is also self-dual, and applying the all-or-none generator to this system yields another lattice with a site threshold of $1/2$, but with non-planar bonds.

\begin{figure}[htbp] 
   \centering
   \includegraphics[width=4in]{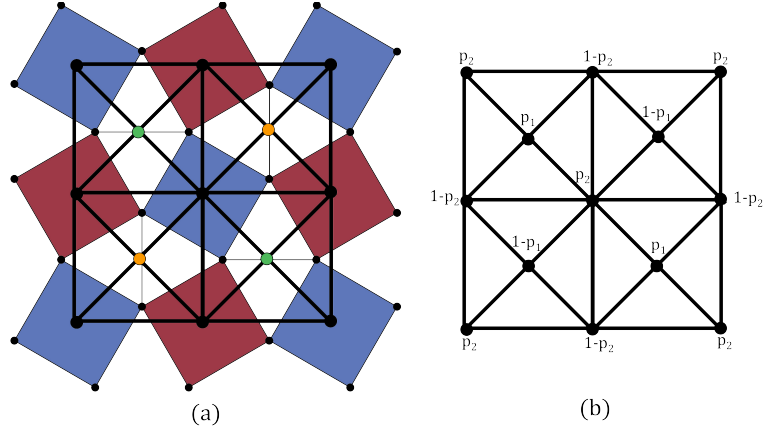} 
   \caption{Hypergraph B and its equivalent site percolation system. (a) We replace the bonds with probability $p_1$ and $1-p_1$ with green and orange sites respectively. More generally, we can also have hyperedges with alternate probability $p_2$  and $1-p_2$, a configuration which we demonstrate by blue and red squares respectively. (b) generalized site percolation on the union-jack lattice with site probabilities $p_1$, $1-p_1$, $p_2$, and $1-p_2$.}
   \label{fig:mostgeneral}
\end{figure}

  Applying the all-or-none generator to hypergraph B, we find site percolation on the union-jack lattice with non-uniform probabilities $p_1, (1-p_1)$ and $1/2$, as shown in Fig.\ \ref{fig:mostgeneral}.   In fact, as shown in that figure, we can generalize this further by having a hypergraph with alternating dual hyperedges, where the blue hyperedge is the dual to the red hyperedge, and the system is still self-dual.  This yields a generalization to a union-jack lattice with probabilities $p_1$ and $p_2$ as shown in Fig.\ \ref{fig:mostgeneral}b.   This is an interesting system that interpolates continuously between the uniform union-jack lattice ($p_1 = p_2 = 1/2$) and the covering lattice or line graph for bond percolation on the square lattice ($p_1 = 0$ or $1$).

 
\subsection{Other lattices}

The finding of an inhomogeneous site percolation model motivated us to look for other soluble cases of inhomogenous site percolation models to compare with.  Here we list several of them.  For comparison, we can think of our union-jack system as having sites with probabilities $p_1$, $p_2$, $p_3$, and $p_4$, with $p_1 + p_3 = 1$ and $p_2 + p_4 = 1$, implying the condition $(1 - p_1 - p_3)(1 - p_2 - p_4) = 0$ or 
\begin{equation}
1 - p_1 - p_2 - p_3 - p_4 + p_1 p_2 + p_1 p_4 + p_2 p_3 + p_3 p_4 = 0
\end{equation}

For inhomogeneous site percolation on the kagome lattice---the covering lattice of the honeycomb lattice---Sykes and Essam \cite{SykesEssam64} showed that the threshold is given by
\begin{equation}
1  - p_1 p_2 - p_1 p_3 - p_2 p_3 + p_1 p_2 p_3 = 0
\end{equation}
In the limit $p_3 \to 1$, this gives $p_1 + p_2 = 1$, in which case the system becomes the non-planar square-lattice covering lattice, similar to what happens to our union-jack lattice when $p_1 \to 0$ or 1.

\begin{figure}[htbp] 
   \centering
   \includegraphics[width=4in]{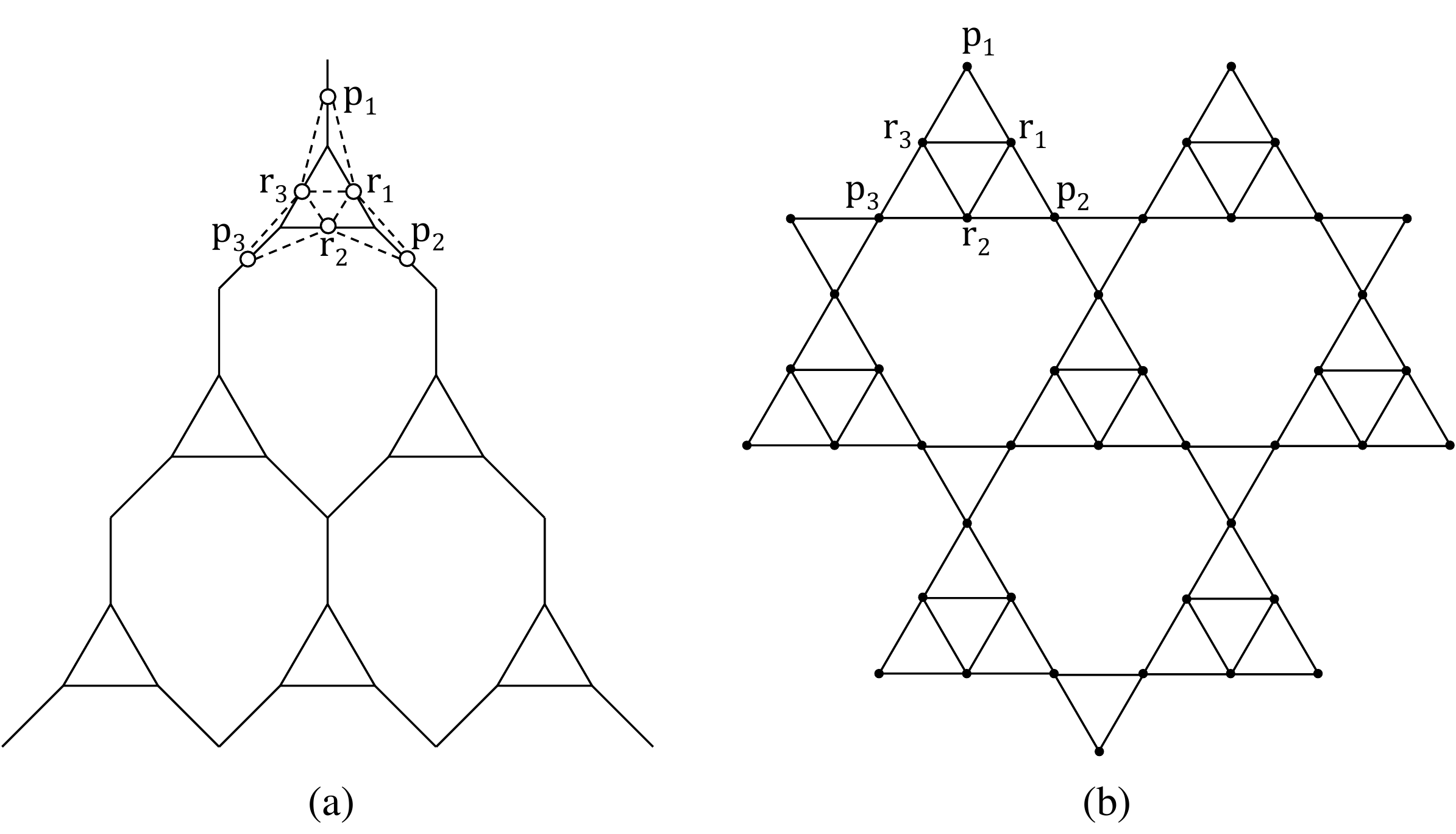} 
   \caption{(a) The inhomogeneous martini lattice. The outer star of bonds of the generator has probabilities $p_1$, $p_2$ and $p_3$, and the inner triangle bonds have probabilities $r_1$, $r_2$ and $r_3$. The covering lattice transformation is also shown on the top generator where each bond is replaced with a site having the same probability. Two sites are connected by a dashed line if the corresponding bonds are connected. (b) the equivalent covering lattice, which is also called the $(1\times1)\space:\space(2\times2)$ subnet lattice \cite{DingFuGuoWu10}.}
   \label{fig:martini}
\end{figure}

We can find the site threshold for the inhomogeneous martini lattice \cite{Scullard06} by considering a martini generator with the three outer bonds having probability $p_1$, $p_2$ and $p_3$, and triangular part correlated with all vertices connected with probability $p_4$ and disconnected with probability $1-p_4$.  The covering lattice to this is the martini site lattice, and using (\ref{eq:allequalsnone}) we find that the threshold is given by
\begin{equation}
1 - p_1 p_2 p_4 - p_1 p_3 p_4 - p_2 p_3 p_4 + p_1 p_2 p_3 p_4 = 0
\end{equation}
For the covering lattice of the inhomogeneous martini lattice, shown in Fig.\ \ref{fig:martini}b, the site threshold  is \cite{Wu06,ziff06,DingFuGuoWu10,Wu10}
\begin{eqnarray}
1&-&p_1 p_2 r_3-p_2 p_3 r_1-p_1 p_3 r_2-p_1 p_2 r_1 r_2-p_1 p_3 r_1 r_3 -p_2 p_3 r_2 r_3  \cr
&+&p_1 p_2 p_3 r_1 r_2 + p_1 p_2 p_3 r_1 r_3 + p_1 p_2 p_3 r_2 r_3
+p_1 p_2 r_1 r_2 r_3 \cr &+& p_1 p_3 r_1 r_2 r_3 + p_2 p_3 r_1 r_2 r_3-2 p_1 p_2 p_3 r_1 r_2 r_3 = 0
\end{eqnarray}   
which interpolates between site percolation on the inhomogeneous triangular covering lattice (by setting all $p$'s equal to 1) and site percolation on the inhomogeneous kagome lattice (setting all $r$'s equal to 1).

While there evidently exists several examples of site thresholds for non-uniform systems, none of these yield the union-jack lattice in any limit, and consequently do not yield our result given above.

\section{Conclusions}
The self-duality and criticality of a hypergraph with four-edges is determined by the duality conditions (\ref{eqn1}), (\ref{eqn2}) and (\ref{eqn3}) given by Bollob\'as and Riordan.  While it is easy to construct generators with correlated bonds that satisfy these conditions, it is not obvious that one can do it with independently occupied bonds, or independent bonds and internal sites.  In this paper we showed that this can indeed be done, making use of three different four-vertex generators (I, II, and III).  We imposed the self-duality conditions and found the percolation thresholds for these three generators, as given in (\ref{eqn:genIIsol}--\ref{eqn:genIIIsol}).  These critical probabilities are independent of the hypergraph, as long as the hypergraph is self-dual.  Then we carried out Monte-Carlo simulations  for generators I and II on the two self-dual hypergraphs (A and B) and observed these critical probabilities indeed give the percolation threshold for the corresponding generator, independent of the irrelevant $p_1$ for hypergraph B. \ Furthermore, we found for generator II on hypergraph A the homogeneous threshold $p = r = t = 0.441374\pm 0.000001$, in agreement with the bounds (\ref{eq:wiermanbounds}) found by Wierman et al.\ \cite{Wierman12}.  Because the uniform generator is not self-dual, this homogeneous threshold is specific to that hypergraph (A), and can result in criticality in hypergraph B only if we choose $p_1\approx0.7665$ or $p_1\approx 0.2335$.  We observe that although for generator I there are  three self-dual critical points, there are also infinitely many points very close to being critical. This can be seen from Fig.\ \ref{fig:contour}, \ref{fig:r-t}  and \ref{fig:P2vP1}. 

Using the all-or-none generator, we can find models for site percolation based upon the hypergraphs.   With this generator on hypergraph B we found a non-uniform union-jack model that interpolates between the uniform union-jack lattice and bond percolation on the square lattice, providing a range of simple fully-triangulated systems that maybe useful for criticality studies.  For example, one can see how the lattice metric factor \cite{HuLinChen95} or the number of clusters per site \cite{ZiffFinchAdamchik97} varies as one goes from one system to the other.

This work shows that while four-hypergraphs are perhaps more restricted than the case of three-hypergraphs (where many self-dual configurations and also uniform-probability generators can easily be found), there are still exact systems with independent bond and site occupancies that can be constructed.   It would  be interesting to find other examples of self-dual four-hypergraphs  and study other generators than the ones considered here, perhaps some that satisfy the duality conditions with two or one distinct probability.

\section*{References}
\bibliography{bibliography}

\vfill\eject
\title[]{Supplementary material for ``Percolation on hypergraphs with four-edges"}

\author{Ojan Khatib Damavandi$^{\,1}$ and Robert M. Ziff$^{\,2}$}

\address{$^1$Department of Physics, University of Michigan, Ann Arbor, MI 48109-1040, USA\\
$^2$Center for the Study of Complex Systems 
and Department of Chemical Engineering, University of Michigan, 
Ann Arbor, MI 48109-2136, USA}
\ead{ojan@umich.edu, rziff@umich.edu}
\vspace{10pt}
\begin{indented}
\item[]August 2015
\end{indented}

\begin{abstract}
Supplementary material.
\end{abstract}

%
%

\maketitle

\section{Exact connection polynomials}

Following are exact enumeration polynomials for $P_1(p,r,t)$ through $P_6(p,r,t)$, defined in Eq.\ (2) of the paper, for generators I, II, and III, where $q=1-p$, $s=1-r$ and $u=1-t$:

\subsection{Generator I}

$P_1 = 4 q^4 r s^3 t^4+q^4 s^4 t^4+16 q^4 r s^3 t^3 u+4 q^4 s^4 t^3 u+20 q^4 r^2 s^2 t^2 u^2+24 q^4 r s^3 t^2 u^2+6 q^4 s^4 t^2 u^2+8 q^4 r^3 s t u^3+20 q^4
   r^2 s^2 t u^3+16 q^4 r s^3 t u^3+4 q^4 s^4 t u^3+q^4 r^4 u^4+4 q^4 r^3 s u^4+6 q^4 r^2 s^2 u^4+4 q^4 r s^3 u^4+q^4 s^4 u^4$ \\\\
   $P_2 = p q^3 r^2 s^2 t^4+q^4 r^2 s^2 t^4+4 p q^3 r s^3 t^4+p q^3 s^4 t^4+4 p q^3 r^2
   s^2 t^3 u+4 q^4 r^2 s^2 t^3 u+16 p q^3 r s^3 t^3 u+4 p q^3 s^4 t^3 u+4 p q^3
   r^3 s t^2 u^2+4 q^4 r^3 s t^2 u^2+23 p q^3 r^2 s^2 t^2 u^2+3 q^4 r^2 s^2 t^2
   u^2+24 p q^3 r s^3 t^2 u^2+6 p q^3 s^4 t^2 u^2+p q^3 r^4 t u^3+q^4 r^4 t
   u^3+10 p q^3 r^3 s t u^3+2 q^4 r^3 s t u^3+21 p q^3 r^2 s^2 t u^3+q^4 r^2 s^2
   t u^3+16 p q^3 r s^3 t u^3+4 p q^3 s^4 t u^3+p q^3 r^4 u^4+4 p q^3 r^3 s
   u^4+6 p q^3 r^2 s^2 u^4+4 p q^3 r s^3 u^4+p q^3 s^4 u^4$\\\\
   $P_3 = p^2 q^2 r^3 s t^4+2 p q^3 r^3 s t^4+q^4 r^3 s t^4+3 p^2 q^2 r^2 s^2 t^4+4 p q^3
   r^2 s^2 t^4+4 p^2 q^2 r s^3 t^4+p^2 q^2 s^4 t^4+4 p^2 q^2 r^3 s t^3 u+8 p q^3
   r^3 s t^3 u+4 q^4 r^3 s t^3 u+12 p^2 q^2 r^2 s^2 t^3 u+16 p q^3 r^2 s^2 t^3
   u+16 p^2 q^2 r s^3 t^3 u+4 p^2 q^2 s^4 t^3 u+p^2 q^2 r^4 t^2 u^2+2 p q^3 r^4
   t^2 u^2+q^4 r^4 t^2 u^2+11 p^2 q^2 r^3 s t^2 u^2+14 p q^3 r^3 s t^2 u^2+q^4
   r^3 s t^2 u^2+28 p^2 q^2 r^2 s^2 t^2 u^2+10 p q^3 r^2 s^2 t^2 u^2+24 p^2 q^2
   r s^3 t^2 u^2+6 p^2 q^2 s^4 t^2 u^2+2 p^2 q^2 r^4 t u^3+2 p q^3 r^4 t u^3+12
   p^2 q^2 r^3 s t u^3+4 p q^3 r^3 s t u^3+22 p^2 q^2 r^2 s^2 t u^3+2 p q^3 r^2
   s^2 t u^3+16 p^2 q^2 r s^3 t u^3+4 p^2 q^2 s^4 t u^3+p^2 q^2 r^4 u^4+4 p^2
   q^2 r^3 s u^4+6 p^2 q^2 r^2 s^2 u^4+4 p^2 q^2 r s^3 u^4+p^2 q^2 s^4 u^4$\\\\
$P_4 = p^4 r^4 t^4+4 p^3 q r^4 t^4+6 p^2 q^2 r^4 t^4+4 p q^3 r^4 t^4+q^4 r^4 t^4+4 p^4 r^3 s t^4+16 p^3 q r^3 s t^4+20 p^2 q^2 r^3 s t^4+8 p q^3 r^3 s t^4+6
   p^4 r^2 s^2 t^4+24 p^3 q r^2 s^2 t^4+20 p^2 q^2 r^2 s^2 t^4+4 p^4 r s^3 t^4+16 p^3 q r s^3 t^4+p^4 s^4 t^4+4 p^3 q s^4 t^4+4 p^4 r^4 t^3 u+16 p^3 q
   r^4 t^3 u+24 p^2 q^2 r^4 t^3 u+16 p q^3 r^4 t^3 u+4 q^4 r^4 t^3 u+16 p^4 r^3 s t^3 u+64 p^3 q r^3 s t^3 u+80 p^2 q^2 r^3 s t^3 u+32 p q^3 r^3 s t^3
   u+24 p^4 r^2 s^2 t^3 u+96 p^3 q r^2 s^2 t^3 u+80 p^2 q^2 r^2 s^2 t^3 u+16 p^4 r s^3 t^3 u+64 p^3 q r s^3 t^3 u+4 p^4 s^4 t^3 u+16 p^3 q s^4 t^3 u+6
   p^4 r^4 t^2 u^2+24 p^3 q r^4 t^2 u^2+30 p^2 q^2 r^4 t^2 u^2+12 p q^3 r^4 t^2 u^2+24 p^4 r^3 s t^2 u^2+96 p^3 q r^3 s t^2 u^2+84 p^2 q^2 r^3 s t^2
   u^2+8 p q^3 r^3 s t^2 u^2+36 p^4 r^2 s^2 t^2 u^2+144 p^3 q r^2 s^2 t^2 u^2+52 p^2 q^2 r^2 s^2 t^2 u^2+24 p^4 r s^3 t^2 u^2+96 p^3 q r s^3 t^2 u^2+6
   p^4 s^4 t^2 u^2+24 p^3 q s^4 t^2 u^2+4 p^4 r^4 t u^3+16 p^3 q r^4 t u^3+12 p^2 q^2 r^4 t u^3+16 p^4 r^3 s t u^3+64 p^3 q r^3 s t u^3+24 p^2 q^2 r^3
   s t u^3+24 p^4 r^2 s^2 t u^3+96 p^3 q r^2 s^2 t u^3+12 p^2 q^2 r^2 s^2 t u^3+16 p^4 r s^3 t u^3+64 p^3 q r s^3 t u^3+4 p^4 s^4 t u^3+16 p^3 q s^4 t
   u^3+p^4 r^4 u^4+4 p^3 q r^4 u^4+4 p^4 r^3 s u^4+16 p^3 q r^3 s u^4+6 p^4 r^2 s^2 u^4+24 p^3 q r^2 s^2 u^4+4 p^4 r s^3 u^4+16 p^3 q r s^3 u^4+p^4
   s^4 u^4+4 p^3 q s^4 u^4$ \\\\
   $P_5 = q^4 r^2 s^2 t^4+4 q^4 r^2 s^2 t^3 u+2 q^4 r^3 s t^2 u^2+2 q^4 r^2 s^2 t^2 u^2$ \\\\
      $P_6 = 2 p^2 q^2 r^2 s^2 t^4+2 p q^3 r^2 s^2 t^4+4 p^2 q^2 r s^3 t^4+p^2 q^2 s^4 t^4+8
   p^2 q^2 r^2 s^2 t^3 u+8 p q^3 r^2 s^2 t^3 u+16 p^2 q^2 r s^3 t^3 u+4 p^2 q^2
   s^4 t^3 u+p^2 q^2 r^4 t^2 u^2+2 p q^3 r^4 t^2 u^2+q^4 r^4 t^2 u^2+8 p^2 q^2
   r^3 s t^2 u^2+8 p q^3 r^3 s t^2 u^2+26 p^2 q^2 r^2 s^2 t^2 u^2+6 p q^3 r^2
   s^2 t^2 u^2+24 p^2 q^2 r s^3 t^2 u^2+6 p^2 q^2 s^4 t^2 u^2+2 p^2 q^2 r^4 t
   u^3+2 p q^3 r^4 t u^3+12 p^2 q^2 r^3 s t u^3+4 p q^3 r^3 s t u^3+22 p^2 q^2
   r^2 s^2 t u^3+2 p q^3 r^2 s^2 t u^3+16 p^2 q^2 r s^3 t u^3+4 p^2 q^2 s^4 t
   u^3+p^2 q^2 r^4 u^4+4 p^2 q^2 r^3 s u^4+6 p^2 q^2 r^2 s^2 u^4+4 p^2 q^2 r s^3
   u^4+p^2 q^2 s^4 u^4$ \\\\
   
 \subsection{Generator II}  

$P_1 = 4 p^2 q^6 r^4 t^4+8 p q^7 r^4 t^4+q^8 r^4 t^4+16 p^2 q^6 r^3 s t^4+32 p q^7 r^3
   s t^4+4 q^8 r^3 s t^4+24 p^2 q^6 r^2 s^2 t^4+48 p q^7 r^2 s^2 t^4+6 q^8 r^2
   s^2 t^4+16 p^2 q^6 r s^3 t^4+32 p q^7 r s^3 t^4+4 q^8 r s^3 t^4+4 p^2 q^6 s^4
   t^4+8 p q^7 s^4 t^4+q^8 s^4 t^4+16 p^2 q^6 r^4 t^3 u+32 p q^7 r^4 t^3 u+4 q^8
   r^4 t^3 u+64 p^2 q^6 r^3 s t^3 u+128 p q^7 r^3 s t^3 u+16 q^8 r^3 s t^3 u+16
   p^3 q^5 r^2 s^2 t^3 u+136 p^2 q^6 r^2 s^2 t^3 u+192 p q^7 r^2 s^2 t^3 u+24
   q^8 r^2 s^2 t^3 u+32 p^3 q^5 r s^3 t^3 u+144 p^2 q^6 r s^3 t^3 u+128 p q^7 r
   s^3 t^3 u+16 q^8 r s^3 t^3 u+16 p^3 q^5 s^4 t^3 u+56 p^2 q^6 s^4 t^3 u+32 p
   q^7 s^4 t^3 u+4 q^8 s^4 t^3 u+24 p^2 q^6 r^4 t^2 u^2+48 p q^7 r^4 t^2 u^2+6
   q^8 r^4 t^2 u^2+96 p^2 q^6 r^3 s t^2 u^2+192 p q^7 r^3 s t^2 u^2+24 q^8 r^3 s
   t^2 u^2+4 p^4 q^4 r^2 s^2 t^2 u^2+80 p^3 q^5 r^2 s^2 t^2 u^2+320 p^2 q^6 r^2
   s^2 t^2 u^2+288 p q^7 r^2 s^2 t^2 u^2+36 q^8 r^2 s^2 t^2 u^2+20 p^4 q^4 r s^3
   t^2 u^2+176 p^3 q^5 r s^3 t^2 u^2+380 p^2 q^6 r s^3 t^2 u^2+192 p q^7 r s^3
   t^2 u^2+24 q^8 r s^3 t^2 u^2+16 p^4 q^4 s^4 t^2 u^2+112 p^3 q^5 s^4 t^2
   u^2+124 p^2 q^6 s^4 t^2 u^2+48 p q^7 s^4 t^2 u^2+6 q^8 s^4 t^2 u^2+16 p^2 q^6
   r^4 t u^3+32 p q^7 r^4 t u^3+4 q^8 r^4 t u^3+64 p^2 q^6 r^3 s t u^3+128 p q^7
   r^3 s t u^3+16 q^8 r^3 s t u^3+8 p^4 q^4 r^2 s^2 t u^3+128 p^3 q^5 r^2 s^2 t
   u^3+368 p^2 q^6 r^2 s^2 t u^3+192 p q^7 r^2 s^2 t u^3+24 q^8 r^2 s^2 t u^3+64
   p^4 q^4 r s^3 t u^3+320 p^3 q^5 r s^3 t u^3+336 p^2 q^6 r s^3 t u^3+128 p q^7
   r s^3 t u^3+16 q^8 r s^3 t u^3+64 p^4 q^4 s^4 t u^3+128 p^3 q^5 s^4 t u^3+96
   p^2 q^6 s^4 t u^3+32 p q^7 s^4 t u^3+4 q^8 s^4 t u^3+4 p^2 q^6 r^4 u^4+8 p
   q^7 r^4 u^4+q^8 r^4 u^4+16 p^2 q^6 r^3 s u^4+32 p q^7 r^3 s u^4+4 q^8 r^3 s
   u^4+2 p^4 q^4 r^2 s^2 u^4+32 p^3 q^5 r^2 s^2 u^4+92 p^2 q^6 r^2 s^2 u^4+48 p
   q^7 r^2 s^2 u^4+6 q^8 r^2 s^2 u^4+16 p^4 q^4 r s^3 u^4+80 p^3 q^5 r s^3
   u^4+84 p^2 q^6 r s^3 u^4+32 p q^7 r s^3 u^4+4 q^8 r s^3 u^4+16 p^4 q^4 s^4
   u^4+32 p^3 q^5 s^4 u^4+24 p^2 q^6 s^4 u^4+8 p q^7 s^4 u^4+q^8 s^4 u^4$ \\\\
$P_2 = p^4 q^4 r^4 t^4+4 p^3 q^5 r^4 t^4+4 p^2 q^6 r^4 t^4+4 p^4 q^4 r^3 s t^4+16 p^3
   q^5 r^3 s t^4+16 p^2 q^6 r^3 s t^4+6 p^4 q^4 r^2 s^2 t^4+24 p^3 q^5 r^2 s^2
   t^4+24 p^2 q^6 r^2 s^2 t^4+4 p^4 q^4 r s^3 t^4+16 p^3 q^5 r s^3 t^4+16 p^2
   q^6 r s^3 t^4+p^4 q^4 s^4 t^4+4 p^3 q^5 s^4 t^4+4 p^2 q^6 s^4 t^4+4 p^4 q^4
   r^4 t^3 u+16 p^3 q^5 r^4 t^3 u+16 p^2 q^6 r^4 t^3 u+16 p^4 q^4 r^3 s t^3 u+64
   p^3 q^5 r^3 s t^3 u+64 p^2 q^6 r^3 s t^3 u+2 p^5 q^3 r^2 s^2 t^3 u+36 p^4 q^4
   r^2 s^2 t^3 u+110 p^3 q^5 r^2 s^2 t^3 u+90 p^2 q^6 r^2 s^2 t^3 u+4 p^5 q^3 r
   s^3 t^3 u+40 p^4 q^4 r s^3 t^3 u+92 p^3 q^5 r s^3 t^3 u+52 p^2 q^6 r s^3 t^3
   u+2 p^5 q^3 s^4 t^3 u+16 p^4 q^4 s^4 t^3 u+30 p^3 q^5 s^4 t^3 u+10 p^2 q^6
   s^4 t^3 u+6 p^4 q^4 r^4 t^2 u^2+24 p^3 q^5 r^4 t^2 u^2+24 p^2 q^6 r^4 t^2
   u^2+24 p^4 q^4 r^3 s t^2 u^2+96 p^3 q^5 r^3 s t^2 u^2+96 p^2 q^6 r^3 s t^2
   u^2+10 p^5 q^3 r^2 s^2 t^2 u^2+92 p^4 q^4 r^2 s^2 t^2 u^2+206 p^3 q^5 r^2 s^2
   t^2 u^2+118 p^2 q^6 r^2 s^2 t^2 u^2+24 p^5 q^3 r s^3 t^2 u^2+130 p^4 q^4 r
   s^3 t^2 u^2+186 p^3 q^5 r s^3 t^2 u^2+54 p^2 q^6 r s^3 t^2 u^2+14 p^5 q^3 s^4
   t^2 u^2+59 p^4 q^4 s^4 t^2 u^2+44 p^3 q^5 s^4 t^2 u^2+9 p^2 q^6 s^4 t^2 u^2+4
   p^4 q^4 r^4 t u^3+16 p^3 q^5 r^4 t u^3+16 p^2 q^6 r^4 t u^3+16 p^4 q^4 r^3 s
   t u^3+64 p^3 q^5 r^3 s t u^3+64 p^2 q^6 r^3 s t u^3+16 p^5 q^3 r^2 s^2 t
   u^3+112 p^4 q^4 r^2 s^2 t u^3+192 p^3 q^5 r^2 s^2 t u^3+56 p^2 q^6 r^2 s^2 t
   u^3+48 p^5 q^3 r s^3 t u^3+168 p^4 q^4 r s^3 t u^3+120 p^3 q^5 r s^3 t u^3+24
   p^2 q^6 r s^3 t u^3+32 p^5 q^3 s^4 t u^3+48 p^4 q^4 s^4 t u^3+24 p^3 q^5 s^4
   t u^3+4 p^2 q^6 s^4 t u^3+p^4 q^4 r^4 u^4+4 p^3 q^5 r^4 u^4+4 p^2 q^6 r^4
   u^4+4 p^4 q^4 r^3 s u^4+16 p^3 q^5 r^3 s u^4+16 p^2 q^6 r^3 s u^4+4 p^5 q^3
   r^2 s^2 u^4+28 p^4 q^4 r^2 s^2 u^4+48 p^3 q^5 r^2 s^2 u^4+14 p^2 q^6 r^2 s^2
   u^4+12 p^5 q^3 r s^3 u^4+42 p^4 q^4 r s^3 u^4+30 p^3 q^5 r s^3 u^4+6 p^2 q^6
   r s^3 u^4+8 p^5 q^3 s^4 u^4+12 p^4 q^4 s^4 u^4+6 p^3 q^5 s^4 u^4+p^2 q^6 s^4
   u^4$\\\\
$P_3 = p^6 q^2 r^4 t^4+6 p^5 q^3 r^4 t^4+12 p^4 q^4 r^4 t^4+8 p^3 q^5 r^4 t^4+4 p^6 q^2
   r^3 s t^4+24 p^5 q^3 r^3 s t^4+48 p^4 q^4 r^3 s t^4+32 p^3 q^5 r^3 s t^4+6
   p^6 q^2 r^2 s^2 t^4+36 p^5 q^3 r^2 s^2 t^4+72 p^4 q^4 r^2 s^2 t^4+48 p^3 q^5
   r^2 s^2 t^4+4 p^6 q^2 r s^3 t^4+24 p^5 q^3 r s^3 t^4+48 p^4 q^4 r s^3 t^4+32
   p^3 q^5 r s^3 t^4+p^6 q^2 s^4 t^4+6 p^5 q^3 s^4 t^4+12 p^4 q^4 s^4 t^4+8 p^3
   q^5 s^4 t^4+4 p^6 q^2 r^4 t^3 u+24 p^5 q^3 r^4 t^3 u+48 p^4 q^4 r^4 t^3 u+32
   p^3 q^5 r^4 t^3 u+16 p^6 q^2 r^3 s t^3 u+96 p^5 q^3 r^3 s t^3 u+192 p^4 q^4
   r^3 s t^3 u+128 p^3 q^5 r^3 s t^3 u+26 p^6 q^2 r^2 s^2 t^3 u+150 p^5 q^3 r^2
   s^2 t^3 u+282 p^4 q^4 r^2 s^2 t^3 u+172 p^3 q^5 r^2 s^2 t^3 u+20 p^6 q^2 r
   s^3 t^3 u+108 p^5 q^3 r s^3 t^3 u+180 p^4 q^4 r s^3 t^3 u+88 p^3 q^5 r s^3
   t^3 u+6 p^6 q^2 s^4 t^3 u+30 p^5 q^3 s^4 t^3 u+42 p^4 q^4 s^4 t^3 u+12 p^3
   q^5 s^4 t^3 u+6 p^6 q^2 r^4 t^2 u^2+36 p^5 q^3 r^4 t^2 u^2+72 p^4 q^4 r^4 t^2
   u^2+48 p^3 q^5 r^4 t^2 u^2+24 p^6 q^2 r^3 s t^2 u^2+144 p^5 q^3 r^3 s t^2
   u^2+288 p^4 q^4 r^3 s t^2 u^2+192 p^3 q^5 r^3 s t^2 u^2+44 p^6 q^2 r^2 s^2
   t^2 u^2+238 p^5 q^3 r^2 s^2 t^2 u^2+400 p^4 q^4 r^2 s^2 t^2 u^2+200 p^3 q^5
   r^2 s^2 t^2 u^2+40 p^6 q^2 r s^3 t^2 u^2+178 p^5 q^3 r s^3 t^2 u^2+218 p^4
   q^4 r s^3 t^2 u^2+54 p^3 q^5 r s^3 t^2 u^2+14 p^6 q^2 s^4 t^2 u^2+48 p^5 q^3
   s^4 t^2 u^2+28 p^4 q^4 s^4 t^2 u^2+4 p^3 q^5 s^4 t^2 u^2+4 p^6 q^2 r^4 t
   u^3+24 p^5 q^3 r^4 t u^3+48 p^4 q^4 r^4 t u^3+32 p^3 q^5 r^4 t u^3+16 p^6 q^2
   r^3 s t u^3+96 p^5 q^3 r^3 s t u^3+192 p^4 q^4 r^3 s t u^3+128 p^3 q^5 r^3 s
   t u^3+36 p^6 q^2 r^2 s^2 t u^3+176 p^5 q^3 r^2 s^2 t u^3+236 p^4 q^4 r^2 s^2
   t u^3+56 p^3 q^5 r^2 s^2 t u^3+40 p^6 q^2 r s^3 t u^3+120 p^5 q^3 r s^3 t
   u^3+64 p^4 q^4 r s^3 t u^3+8 p^3 q^5 r s^3 t u^3+16 p^6 q^2 s^4 t u^3+16 p^5
   q^3 s^4 t u^3+4 p^4 q^4 s^4 t u^3+p^6 q^2 r^4 u^4+6 p^5 q^3 r^4 u^4+12 p^4
   q^4 r^4 u^4+8 p^3 q^5 r^4 u^4+4 p^6 q^2 r^3 s u^4+24 p^5 q^3 r^3 s u^4+48 p^4
   q^4 r^3 s u^4+32 p^3 q^5 r^3 s u^4+9 p^6 q^2 r^2 s^2 u^4+44 p^5 q^3 r^2 s^2
   u^4+59 p^4 q^4 r^2 s^2 u^4+14 p^3 q^5 r^2 s^2 u^4+10 p^6 q^2 r s^3 u^4+30 p^5
   q^3 r s^3 u^4+16 p^4 q^4 r s^3 u^4+2 p^3 q^5 r s^3 u^4+4 p^6 q^2 s^4 u^4+4
   p^5 q^3 s^4 u^4+p^4 q^4 s^4 u^4$\\\\
$P_4 = p^8 r^4 t^4+8 p^7 q r^4 t^4+24 p^6 q^2 r^4 t^4+32 p^5 q^3 r^4 t^4+16 p^4 q^4 r^4
   t^4+4 p^8 r^3 s t^4+32 p^7 q r^3 s t^4+96 p^6 q^2 r^3 s t^4+128 p^5 q^3 r^3 s
   t^4+64 p^4 q^4 r^3 s t^4+6 p^8 r^2 s^2 t^4+48 p^7 q r^2 s^2 t^4+144 p^6 q^2
   r^2 s^2 t^4+192 p^5 q^3 r^2 s^2 t^4+96 p^4 q^4 r^2 s^2 t^4+4 p^8 r s^3 t^4+32
   p^7 q r s^3 t^4+96 p^6 q^2 r s^3 t^4+128 p^5 q^3 r s^3 t^4+64 p^4 q^4 r s^3
   t^4+p^8 s^4 t^4+8 p^7 q s^4 t^4+24 p^6 q^2 s^4 t^4+32 p^5 q^3 s^4 t^4+16 p^4
   q^4 s^4 t^4+4 p^8 r^4 t^3 u+32 p^7 q r^4 t^3 u+96 p^6 q^2 r^4 t^3 u+128 p^5
   q^3 r^4 t^3 u+64 p^4 q^4 r^4 t^3 u+16 p^8 r^3 s t^3 u+128 p^7 q r^3 s t^3
   u+384 p^6 q^2 r^3 s t^3 u+512 p^5 q^3 r^3 s t^3 u+256 p^4 q^4 r^3 s t^3 u+24
   p^8 r^2 s^2 t^3 u+192 p^7 q r^2 s^2 t^3 u+564 p^6 q^2 r^2 s^2 t^3 u+720 p^5
   q^3 r^2 s^2 t^3 u+336 p^4 q^4 r^2 s^2 t^3 u+16 p^8 r s^3 t^3 u+128 p^7 q r
   s^3 t^3 u+360 p^6 q^2 r s^3 t^3 u+416 p^5 q^3 r s^3 t^3 u+160 p^4 q^4 r s^3
   t^3 u+4 p^8 s^4 t^3 u+32 p^7 q s^4 t^3 u+84 p^6 q^2 s^4 t^3 u+80 p^5 q^3 s^4
   t^3 u+16 p^4 q^4 s^4 t^3 u+6 p^8 r^4 t^2 u^2+48 p^7 q r^4 t^2 u^2+144 p^6 q^2
   r^4 t^2 u^2+192 p^5 q^3 r^4 t^2 u^2+96 p^4 q^4 r^4 t^2 u^2+24 p^8 r^3 s t^2
   u^2+192 p^7 q r^3 s t^2 u^2+576 p^6 q^2 r^3 s t^2 u^2+768 p^5 q^3 r^3 s t^2
   u^2+384 p^4 q^4 r^3 s t^2 u^2+36 p^8 r^2 s^2 t^2 u^2+288 p^7 q r^2 s^2 t^2
   u^2+812 p^6 q^2 r^2 s^2 t^2 u^2+944 p^5 q^3 r^2 s^2 t^2 u^2+368 p^4 q^4 r^2
   s^2 t^2 u^2+24 p^8 r s^3 t^2 u^2+192 p^7 q r s^3 t^2 u^2+472 p^6 q^2 r s^3
   t^2 u^2+400 p^5 q^3 r s^3 t^2 u^2+64 p^4 q^4 r s^3 t^2 u^2+6 p^8 s^4 t^2
   u^2+48 p^7 q s^4 t^2 u^2+92 p^6 q^2 s^4 t^2 u^2+32 p^5 q^3 s^4 t^2 u^2+2 p^4
   q^4 s^4 t^2 u^2+4 p^8 r^4 t u^3+32 p^7 q r^4 t u^3+96 p^6 q^2 r^4 t u^3+128
   p^5 q^3 r^4 t u^3+64 p^4 q^4 r^4 t u^3+16 p^8 r^3 s t u^3+128 p^7 q r^3 s t
   u^3+384 p^6 q^2 r^3 s t u^3+512 p^5 q^3 r^3 s t u^3+256 p^4 q^4 r^3 s t
   u^3+24 p^8 r^2 s^2 t u^3+192 p^7 q r^2 s^2 t u^3+496 p^6 q^2 r^2 s^2 t
   u^3+448 p^5 q^3 r^2 s^2 t u^3+64 p^4 q^4 r^2 s^2 t u^3+16 p^8 r s^3 t u^3+128
   p^7 q r s^3 t u^3+224 p^6 q^2 r s^3 t u^3+64 p^5 q^3 r s^3 t u^3+4 p^8 s^4 t
   u^3+32 p^7 q s^4 t u^3+16 p^6 q^2 s^4 t u^3+p^8 r^4 u^4+8 p^7 q r^4 u^4+24
   p^6 q^2 r^4 u^4+32 p^5 q^3 r^4 u^4+16 p^4 q^4 r^4 u^4+4 p^8 r^3 s u^4+32 p^7
   q r^3 s u^4+96 p^6 q^2 r^3 s u^4+128 p^5 q^3 r^3 s u^4+64 p^4 q^4 r^3 s u^4+6
   p^8 r^2 s^2 u^4+48 p^7 q r^2 s^2 u^4+124 p^6 q^2 r^2 s^2 u^4+112 p^5 q^3 r^2
   s^2 u^4+16 p^4 q^4 r^2 s^2 u^4+4 p^8 r s^3 u^4+32 p^7 q r s^3 u^4+56 p^6 q^2
   r s^3 u^4+16 p^5 q^3 r s^3 u^4+p^8 s^4 u^4+8 p^7 q s^4 u^4+4 p^6 q^2 s^4 u^4$ \\\\
   $P_5 = p^4 q^4 r^4 t^4+4 p^3 q^5 r^4 t^4+4 p^2 q^6 r^4 t^4+4 p^4 q^4 r^3 s t^4+16 p^3
   q^5 r^3 s t^4+16 p^2 q^6 r^3 s t^4+6 p^4 q^4 r^2 s^2 t^4+24 p^3 q^5 r^2 s^2
   t^4+24 p^2 q^6 r^2 s^2 t^4+4 p^4 q^4 r s^3 t^4+16 p^3 q^5 r s^3 t^4+16 p^2
   q^6 r s^3 t^4+p^4 q^4 s^4 t^4+4 p^3 q^5 s^4 t^4+4 p^2 q^6 s^4 t^4+4 p^4 q^4
   r^4 t^3 u+16 p^3 q^5 r^4 t^3 u+16 p^2 q^6 r^4 t^3 u+16 p^4 q^4 r^3 s t^3 u+64
   p^3 q^5 r^3 s t^3 u+64 p^2 q^6 r^3 s t^3 u+28 p^4 q^4 r^2 s^2 t^3 u+100 p^3
   q^5 r^2 s^2 t^3 u+88 p^2 q^6 r^2 s^2 t^3 u+24 p^4 q^4 r s^3 t^3 u+72 p^3 q^5
   r s^3 t^3 u+48 p^2 q^6 r s^3 t^3 u+8 p^4 q^4 s^4 t^3 u+20 p^3 q^5 s^4 t^3 u+8
   p^2 q^6 s^4 t^3 u+6 p^4 q^4 r^4 t^2 u^2+24 p^3 q^5 r^4 t^2 u^2+24 p^2 q^6 r^4
   t^2 u^2+24 p^4 q^4 r^3 s t^2 u^2+96 p^3 q^5 r^3 s t^2 u^2+96 p^2 q^6 r^3 s
   t^2 u^2+50 p^4 q^4 r^2 s^2 t^2 u^2+156 p^3 q^5 r^2 s^2 t^2 u^2+108 p^2 q^6
   r^2 s^2 t^2 u^2+46 p^4 q^4 r s^3 t^2 u^2+104 p^3 q^5 r s^3 t^2 u^2+38 p^2 q^6
   r s^3 t^2 u^2+16 p^4 q^4 s^4 t^2 u^2+16 p^3 q^5 s^4 t^2 u^2+4 p^2 q^6 s^4 t^2
   u^2+4 p^4 q^4 r^4 t u^3+16 p^3 q^5 r^4 t u^3+16 p^2 q^6 r^4 t u^3+16 p^4 q^4
   r^3 s t u^3+64 p^3 q^5 r^3 s t u^3+64 p^2 q^6 r^3 s t u^3+44 p^4 q^4 r^2 s^2
   t u^3+112 p^3 q^5 r^2 s^2 t u^3+40 p^2 q^6 r^2 s^2 t u^3+32 p^4 q^4 r s^3 t
   u^3+32 p^3 q^5 r s^3 t u^3+8 p^2 q^6 r s^3 t u^3+p^4 q^4 r^4 u^4+4 p^3 q^5
   r^4 u^4+4 p^2 q^6 r^4 u^4+4 p^4 q^4 r^3 s u^4+16 p^3 q^5 r^3 s u^4+16 p^2 q^6
   r^3 s u^4+11 p^4 q^4 r^2 s^2 u^4+28 p^3 q^5 r^2 s^2 u^4+10 p^2 q^6 r^2 s^2
   u^4+8 p^4 q^4 r s^3 u^4+8 p^3 q^5 r s^3 u^4+2 p^2 q^6 r s^3 u^4$ \\\\
   $P_6 = 2 p^6 q^2 r^2 s^2 t^3 u+8 p^5 q^3 r^2 s^2 t^3 u+8 p^4 q^4 r^2 s^2 t^3 u+4 p^6
   q^2 r s^3 t^3 u+16 p^5 q^3 r s^3 t^3 u+16 p^4 q^4 r s^3 t^3 u+2 p^6 q^2 s^4
   t^3 u+8 p^5 q^3 s^4 t^3 u+8 p^4 q^4 s^4 t^3 u+10 p^6 q^2 r^2 s^2 t^2 u^2+40
   p^5 q^3 r^2 s^2 t^2 u^2+40 p^4 q^4 r^2 s^2 t^2 u^2+20 p^6 q^2 r s^3 t^2
   u^2+68 p^5 q^3 r s^3 t^2 u^2+56 p^4 q^4 r s^3 t^2 u^2+10 p^6 q^2 s^4 t^2
   u^2+28 p^5 q^3 s^4 t^2 u^2+11 p^4 q^4 s^4 t^2 u^2+16 p^6 q^2 r^2 s^2 t u^3+64
   p^5 q^3 r^2 s^2 t u^3+64 p^4 q^4 r^2 s^2 t u^3+32 p^6 q^2 r s^3 t u^3+80 p^5
   q^3 r s^3 t u^3+32 p^4 q^4 r s^3 t u^3+16 p^6 q^2 s^4 t u^3+16 p^5 q^3 s^4 t
   u^3+4 p^4 q^4 s^4 t u^3+4 p^6 q^2 r^2 s^2 u^4+16 p^5 q^3 r^2 s^2 u^4+16 p^4
   q^4 r^2 s^2 u^4+8 p^6 q^2 r s^3 u^4+20 p^5 q^3 r s^3 u^4+8 p^4 q^4 r s^3
   u^4+4 p^6 q^2 s^4 u^4+4 p^5 q^3 s^4 u^4+p^4 q^4 s^4 u^4$ \\\\
   
   \subsection{Generator III}
   
   For generator III, the probabilities can be found by hand by going through the different configurations of the outside bonds first, resulting in the first entry.  The second entry is the expansion which was verified by exact enumeration: \\\\
   $P_1=q^4 \left(t \left(4 r s^3+s^4\right)+u\right) = 4 q^4 r s^3 t  +  q^4 s^4 t  +  q^4 u$\\\\
   $P_2=p q^3 \left(t \left(r^2 s^2+4 r
   s^3+s^4\right)+u\right)+q^4 r^2 s^2 t = p q^3 r^2 s^2 t  +  q^4 r^2 s^2 t  +  4 p q^3 r s^3 t  +  p q^3 s^4 t  +  
 p q^3 u$\\\\
   $P_3=p^2 q^2 \left(t \left(r s^3+s\right)+u\right)+2 p q^3
   r s \left(1-s^2\right) t+q^4 r^3 s t = p^2 q^2 r^3 s t + 2 p q^3 r^3 s t + q^4 r^3 s t + 
 3 p^2 q^2 r^2 s^2 t + 4 p q^3 r^2 s^2 t + 4 p^2 q^2 r s^3 t + 
 p^2 q^2 s^4 t + p^2 q^2 u$\\\\
   $P_4=p^4+4 p^3 q+4 p^2 q^2 r \left(1-s^3\right) t+2 p^2 q^2
   \left(1-s^2\right)^2 t+4 p q^3 r^2
   \left(1-s^2\right) t+q^4 r^4 t = p^4 r^4 t + 4 p^3 q r^4 t + 6 p^2 q^2 r^4 t + 4 p q^3 r^4 t + 
 q^4 r^4 t + 4 p^4 r^3 s t + 16 p^3 q r^3 s t + 20 p^2 q^2 r^3 s t + 
 8 p q^3 r^3 s t + 6 p^4 r^2 s^2 t + 24 p^3 q r^2 s^2 t + 
 20 p^2 q^2 r^2 s^2 t + 4 p^4 r s^3 t + 16 p^3 q r s^3 t + 
 p^4 s^4 t + 4 p^3 q s^4 t + p^4 u + 4 p^3 q u$\\\\
   $P_5=q^4 r^2 s^2 t$\\\\
   $P_6=p^2 q^2 \left(t \left(2 r^2 s^2+4 r
   s^3+s^4\right)+u\right)+2 p q^3 r^2 s^2 t = 2 p^2 q^2 r^2 s^2 t + 2 p q^3 r^2 s^2 t + 4 p^2 q^2 r s^3 t + 
 p^2 q^2 s^4 t + p^2 q^2 u$\\\\

\end{document}